\documentclass[12pt,preprint]{aastex}

\shorttitle{Integral Field Spectroscopy of HH 34}
\shortauthors{Beck et al.}

\begin{document}


\title{The Structure of the Inner HH 34 Jet from Optical Integral Field Spectroscopy\altaffilmark{1}}


\author{Tracy L. Beck\altaffilmark{2}, A. Riera\altaffilmark{3,4},  A. C. Raga\altaffilmark{5}, Bo Reipurth\altaffilmark{6}}
\email{tbeck@gemini.edu, angels.riera@upc.es, raga@nucleares.unam.mx, reipurth@ifa.hawaii.edu}


\altaffiltext{1}{Based on observations obtained at the Gemini Observatory, which is operated by the Association of Universities for Research in Astronomy, Inc., under a cooperative agreement with the NSF on behalf of the Gemini partnership: the National Science Foundation (United States), the Particle Physics and Astronomy Research Council (United Kingdom), the National Research Council (Canada), CONICYT (Chile), the Australian Research Council (Australia), CNPq (Brazil), and CONICET (Argentina).}

\altaffiltext{2}{Gemini Observatory, Northern Operations, 670 N. A'ohoku Pl., Hilo, HI, 96720}
\altaffiltext{3}{Departament de F\'\i sica i Enginyeria Nuclear, Universitat Polit\'ecnica de Catalunya, Av. V\'\i ctor 
Balaguers/n E-08800 Vilanova i La Geltr\'u, Spain}
\altaffiltext{4}{Departament d'Astronomia i Meteorologia, Universitat de Barcelona, Av. Diagonal 647, E-08028 Barcelona, Spain}
\altaffiltext{5}{Instituto de Ciencias Nucleares, UNAM, Ap. 70-543, 04510 D.F., M\'exico}
\altaffiltext{6}{Institute for Astronomy, University of Hawaii, 640 N. A'ohoku Pl. Hilo, HI 96720}


\begin{abstract}
  We present high spatial resolution optical integral field spectroscopy of a collimated Herbig-Haro jet viewed nearly edge-on.  Maps of the line emission, velocity centroid, and velocity dispersion were generated for the H$\alpha$ and [S II] emission features from the inner collimated jet and exciting source region of the HH 34 outflow.  The kinematic structure of the jet shows several maxima and minima in both velocity centroid value and velocity dispersion along the jet axis.  Perpendicular to the flow direction the velocity decreases outward from the axis to the limb of the jet, but the velocity dispersion increases.  Maps of the electron density structure were derived from the line ratio of [S~II] 6731/6716 emission.  We have found that the jet exhibits a pronounced ``striped'' pattern in electron density; the high $n_e$ regions are at the leading side of each of the emission knots in the collimated jet, and low $n_e$ regions in the down-flow direction.  On average, the measured electron density decreases outward from the inner regions of the jet, but the highest $n_e$ found in the outflow is spatially offset from the nominal position of the exciting star.  The results of our high spatial resolution optical integral field spectroscopy show very good agreement with the kinematics and electron density structure predicted by the existing internal working surface models of the HH~34 outflow.

\end{abstract}


\keywords{ISM: Herbig-Haro objects -- ISM: jets and outflows --
ISM: kinematics and dynamics -- ISM: individual (HH 34)}


\section{Introduction}

Herbig-Haro (HH) jets represent an important physical mechanism by which newborn stars are able to remove angular momentum from the infalling circumstellar material that builds them (for a review see Reipurth \& Bally 2001).  Such jets are highly supersonic, and the velocity variability within the flows cause internal shocks that are seen as knots within the jet channels.  It is these shocks that produce the characteristic emission line spectra of HH jets (Raga et al. 1990).  High spatial resolution imaging observations of the Hubble Space Telescope (HST) and Wide Field Planetary Camera-2 (WFPC2) have provided detailed insights into the knot morphology and kinematics of many HH jets (e. g.  Heathcote et al. 1996; Ray et al. 1996; Reipurth et al. 1997).  Multi-epoch proper motion studies using WFPC2 have resulted in accurate tangential velocity measurements of several jets (Reipurth et al. 2002).  Some HST studies of the radial velocities of HH jets have been done using the Space Telescope Imaging Spectrograph (STIS; Bacciotti et al. 2002; Raga et al. 2002a; Coffey et al. 2004; Woitas et al. 2005), but the number of high spatial and spectral resolution observations of the velocity structure in HH jets are few.

Several observational studies have used the technique of stepping a spectral slit across a HH jet to obtain spectroscopy with 2D spatial resolution on the sky (Solf et al. 1991; Solf \& B\"ohm 1999; Riera et al. 2001; Bacciotti et al. 2002; Riera et al. 2003a; Coffey et al. 2004; Woitas et al. 2005).  In only one pointing of a telescope, imaging spectroscopy obtained with Integral Field Units (IFUs) can provide three dimensional, x,y,$\lambda$ datacubes at high spatial resolution with simultaneous coverage of many emission lines of interest (as opposed to Fabry-Perot interferometry, which can obtain information over just one feature).   Observations of spatially resolved emission features provide the ability to determine radial velocity structure and line ratio maps from any spatial position within the image cube  (thus excitation and/or electron density structure).  This technique has been applied to a few HH flows with excellent results (Lavalley et al. 1997; Beck et al. 2004; Vasconcelos et al. 2005).

HH 34 is arguably one of the most spectacular of the known HH flows.  It has a well collimated jet consisting of $\sim$20 knots in its inner blueshifted region, and bow shocks placed symmetrically around the exciting source (Reipurth et al. 1986;  B\"uhrke, Mundt, Ray 1988).  The source is a $\sim$28 L$_{\odot}$ young star that is still partly embedded within its natal material (Reipurth et al. 1993).  It has been detected at radio wavelengths, drives a weak molecular outflow, and is surrounded by remaining molecular cloud material (Rodr\'\i guez \& Reipurth 1996; Chernin \& Masson 1995; Stapelfeldt \& Scoville 1993; Davis \& Dent 1993; Anglada et al. 1995).  The HH 34 outflow complex is at a distance of $\sim$450 parsecs.  The HH 34 collimated jet represents the innermost part of a large parsec-scale HH flow that has giant bow shocks in both of its lobes (Bally \& Devine 1994; Devine et al. 1997;  Eisl$\ddot{o}$ffel \& Mundt 1997).  The collimated jet appears to be viewed nearly edge-on, it has an inclination of roughly 30$^{\circ}$ with respect to the plane of the sky (Heathcote \& Reipurth 1992).  The jet itself has been observed with HST imaging in the optical and infrared (Ray et al. 1996; Reipurth et al. 2002; Reipurth et al. 2000).  The proper motion and tangential velocities of the jet knots have been determined from these high spatial resolution studies.  The morphological structure and velocity variability observed in HH 34 on small and large scales has been modeled successfully by Raga \& Noriega-Crespo (1998), Cabrit \& Raga (2000), Masciadri et al. (2002a) and Raga et al. (2002c).

The theoretical models have been good at describing the existing observations of HH 34, but high spatial resolution maps of the radial velocity and electron density structure have never been available for this source.  In the hope that the nearly edge-on viewing angle of this object coupled with high spatial and spectral resolution data could provide new insights into the excitation and density structure of young stellar outflows, we used optical integral field spectroscopy to observe the inner regions of the blueshifted HH~34 jet.  In this paper we present the details of our observations (\S 2) and the results and interpretations for the collimated jet (\S3) and the exciting source (\S4).  In \S5 we summarize the key results from this study.

\section{Observations}

The data for this project were obtained over four nights in 2003 February with the Gemini Multi-Object Spectrograph Integral Field Unit (GMOS IFU; Hook et al. 2003; Allington-Smith et al. 2002) at the Gemini North Fredrick C. Gillett Telescope.  We obtained IFU spectral imaging data on three fields along the HH 34 outflow, one position was at the exciting source of the jet, and two pointings were centered along the collimated jet.  Figure 1 presents an H$\alpha$ image of the HH 34 region that was made from coadding six 5 minute exposures obtained to set up the IFU at our target positions.  The zoomed view on the right side of Figure 1 shows an expanded view of the HH 34 outflow and exciting source, overplotted are the positions of the three IFU fields.  Observations for this program were obtained in photometric conditions.   The nearby star V801 Ori was used to provide guiding, tip-tilt and low-order astigmatism corrections using the GMOS on-instrument wavefront sensor (OIWFS).  The (OIWFS enhanced) seeing for data at the position of the HH 34 exciting source was $\sim0\farcs5$, and the two fields that were centered on the collimated jet were observed in seeing that was $\sim0\farcs65$.  

Our observations were obtained using the IFU in 2-slit mode which provides spectroscopy over a spatial field of 7$''$ x 5$''$ for each pointing.  With this observing configuration, the GMOS IFU is comprised of 1500 fibers; each spans a $0\farcs2$ hexagonal region on the sky.  1000 fibers make up the  7$''$ x 5$''$ science field of view, and 500 fibers make up a smaller, dedicated sky field which is fixed at a 1$'$ distance from the science position (Allington-Smith et al. 2002).   We used the R831 grating in GMOS and prior to sky subtraction in the final data frames an instrumental velocity resolution of $\approx$19.8$\pm$3.4~km~s$^{-1}$ was measured from Gaussian fits to several sky emission lines in the IFU data.  We were able to obtain spectra over the 6220 - 6840\AA\ region, which includes [O~I]~6300/63, [N~II]~6548/84, H$\alpha$, and [S~II]~6716/31 lines.  The r$'$ and RG610 filters were used to limit the overlap of spectra that results from using the IFU in the two-slit mode configuration with a high spectral resolution setting.  Unfortunately, because of the overlap of spectra from the two slits, the [O~I] features were contaminated and we were unable to extract accurate scientific information from these emission lines.

A total of four 1800 second frames were obtained on each of the two pointings on the collimated jet.  We observed two iterations through a 2-position dither pattern with $0\farcs5$ offsets, and the total integration time on each of the two jet positions was 2 hours.  The frames taken at the position of the outflow source were obtained by doing four iterations of a similar two point dither pattern, for a total integration time of 4 hours.  The GMOS IFU scripts available in the Gemini IRAF package were used for reduction of the data, the data analysis discussed in Beck et al. (2004) provides a step-by-step demonstration of the method we employed to extract, wavelength calibrate and reformat the data into a 3-dimensional, x,y,$\lambda$ datacube.   The spatial axes in the 3D datacube were resampled to square pixels with a 0.$''$1 resolution.  We subtracted the nebular emission using the fibers in the dedicated sky field.  At the positions centered on the collimated jet, we also corrected for structure in the nebular component by subtracting a residual gradient in the emission.  The data were mosaiced together by manually offsetting and coadding the individual datacubes using the IDL programming language.  We used the H$\alpha$ image obtained for acquisition of the IFU fields as a guide for calculating accurate offsets to merge the datacubes into two that have larger spatial extent.  The two pointings on the collimated outflow were merged to form a single datacube of the jet (see Figure 1).  





\section{Results: The HH 34 Collimated Jet}

The region of the HH 34 jet observed with the GMOS IFU is shown in Figure 2.  Figure 2 displays a map of the jet in the [S~II] 6716 \AA\ emission that was constructed by summing the flux from the feature over its wavelength extent in the final mosaiced datacube.  The observed region includes six bright knots in the HH 34 collimated jet (from knot E to knot J).  We have identified the observed knots following the nomenclature of B\"uhrke, Mundt \& Ray (1988).  

From the IFU data we have simulated a ``longslit spectrum'' of HH 34 by summing the flux from the HH 34 jet over 1$\farcs$2 on either side of the axis of the outflow.  In Figure 3 we show the position-velocity (P-V) diagrams of the combined [S~II] emission (6716+6731 \AA) and H$\alpha$ features for the data obtained on the collimated jet and the exciting source.  These P-V diagrams are qualitatively similar to those presented in  B\"uhrke, Mundt \& Ray (1988) and Heathcote \& Reipurth (1992).  Panels c and d in Figure 3 present the average line profiles of the [S~II] 6716 \AA\ and H$\alpha$ emission features from the G knot in the HH~34 complex that we derived from the IFU data.  The H$\alpha$ line profile is qualitatively similar to the profile presented by B\"uhrke, Mundt \& Ray (1988); it shows a double-peaked structure with one component at v$_r\approx-130$~km~s$^{-1}$ and one at $\sim-55$~km~s$^{-1}$.  Our data have the spectral resolution and signal-to-noise necessary to detect the [S~II] emission peak from the ``slow'' component seen by B\"uhrke, Mundt \& Ray (1988) and Heathcote \& Reipurth (1992).  However, we find that the [S~II] 6716 \AA\ profile in all of the E - J knots shows only a single emission peak with no obvious second component.

\subsection{Radial Velocity Channel Maps and Velocity Dispersion} 

Figures 4 and 5 present velocity channel maps of the H$\alpha$ and [S~II] 6716 \AA\ emission that were derived from the IFU image cube of the collimated jet.  In the collimated jet, the H$\alpha$ emission is detected at heliocentric radial velocities ranging from low values of $\sim$0~km~s$^{-1}$ to blueshifted velocities of $\sim-160$~km~s$^{-1}$ (Figure 4).  The surface brightness of all of the observed knots increases from low radial velocities until it reaches a maximum emission level at a heliocentric radial velocity from $-$96 to $-$80~km~s$^{-1}$.  At radial velocites between $-$50~km~s$^{-1}$ and $-$5~km~s$^{-1}$, the H$\alpha$ surface brightness of the knots is still relatively high in comparison to the emission of [SII] at comparable radial velocities (Figure 5).  The H$\alpha$ emission at heliocentric radial velocities $v_r$ $\sim$ $-$10~km~s$^{-1}$ to 20~km~s$^{-1}$ may be affected by imperfectly subtracted residual flux from the known nebular emission at $v_r$ $\approx$ +10~km~s$^{-1}$ (Heathcote \& Reipurth 1992).  Figure 5 shows that the [S~II] emission is detected at heliocentric radial velocities ranging from $\sim$ $-$180~km~s$^{-1}$ to $-$15~km~s$^{-1}$.    The surface brightness of the knots in  [S~II] emission grows with increasing radial velocity, reaching a maximum emission level at $-$105 $\rightarrow$ $-$120~km~s$^{-1}$.  

The velocity centroid and velocity dispersion of the emission line profiles are often used to derive more detailed information on the kinematics of HH jets (Riera et al. 2003b).  We have derived the barycentric velocity from the H$\alpha$ and [S~II] 6716 \AA\ emission line profiles at each spatial position in our IFU datacube (Figure 6).   From this analysis we find that the [S~II] heliocentric radial velocity has an approximately constant value of $-$105~km~s$^{-1}$, along most of the central axis of the HH 34 jet.  The most strongly blueshifted radial velocity of $\sim$ $-$116~km~s$^{-1}$ is found at the central position of knot E.  At larger distances from the source (knots F, G and J) the radial velocity along the central axis remains approximately constant, with values oscillating between  $-$105 and $-$110~km~s$^{-1}$. 

The H$\alpha$ heliocentric radial velocity (Figure 6) has a constant velocity along the central axis of the HH 34 jet, with a mean value of $-$85~km~s$^{-1}$, this is similar to the velocity of the ``fast'' component detected by B\"urke et al. (1988).  However, it is worthwhile to note that the velocity barycenter is a measurement of the weighted centroid of the profile and information from multiple emission components is not well represented.  From Figure 3 we found that the H$\alpha$ emission profile is double-peaked in many regions.  This profile is shifting the value of the velocity barycenter toward the slower emission component and thus the H$\alpha$ and [S~II] 6716 \AA\ emission lines have slightly different velocity centroids (by $\approx$20~km~s$^{-1}$).

Along the HH~34 jet axis, both the H$\alpha$ and the [S~II]~6716~\AA\ line barycenter velocities show a series of velocity minima; the velocity in each of the inter-knot regions is slightly less negative than at the positions of the knots.  The most significant difference is seen in the velocity map for [S~II]~6716~\AA\ in the knot E-F region; the velocity in the E knot is $v_r$$\sim-115$~km~s$^{-1}$ and the interknot region has $v_r$$\sim-100$~km~s$^{-1}$.   Over the jet region sampled, the average difference between the maxima and minima of the barycenter velocity along the jet axis is $\Delta v_{r,||}\approx 5$~km~s$^{-1}$. Deprojecting for an orientation angle of $30^\circ$ (between the jet axis and the plane of the sky, see Heathcote \& Reipurth 1992), the intrinsic difference is $\Delta v_{||}\approx 10$~km~s$^{-1}$ between the peaks and valleys of the axial barycenter velocity.

Perpendicular to the jet axis, both the H$\alpha$ and the [S~II]~6716~\AA\ line barycenters show a decrease to less blueshifted velocities.  The difference between the axial velocities and the velocities at the outer edge of the detected jet beam are $\Delta v_{r,\perp}\approx 20$~km~s$^{-1}$ (corresponding to a deprojected $\Delta v_\perp\approx 40$~km~s$^{-1}$). Similar ``center-to-limb'' velocity drops are seen both for jet cross sections taken through the peaks of the knots as well as for cuts across the inter-knot regions. These velocity drops occur over distances of $\sim 1''$ from the jet axis (corresponding to $\sim 450$~AU at the distance of HH~34).  The velocity structure is similar on each side of the outflow axis, we have found no evidence for systematic asymmetries in the barycenter velocities along the cross-jet direction, though some individual knots do show side-to-side barycenter velocity asymmetries (e.g. Figure 6).

Figure 7 shows the H$\alpha$ and [S~II]~6716~\AA\ velocity dispersion maps of the collimated jet.  The H$\alpha$ line is generally broader than the [S~II]~6716~\AA\ line, but the dispersions of both lines have similar properties.  The velocity width shows several maxima and minima along the axis of the jet and moving away from the exciting source (which is off the panel to the right in Figure 7).  The dispersion minima coincide with the positions of the knots, and the maxima with the inter-knot regions. The difference in velocity dispersions along the jet axis is $\Delta\sigma\sim 5$~km~s$^{-1}$ (similar to the corresponding variations in the barycenter velocity, see above).  From Figure 7, we have also found that the velocity dispersion typically grows away from the central axis of the HH~34 jet.  The emission profile cross sections have a $\Delta\sigma \sim 10$~km~s$^{-1}$ increase from the axis to the limb of the detected jet ($\sim 1''$ distant). 

\subsection{Emission Line Ratios}

The integrated intensity maps of the [S~II] 6716, 6731 \AA\ and H$\alpha$ emission lines were constructed by summing each line in the data cube over its wavelength extent.  From these, we have made [S~II] (6716+6731)/H$\alpha$ and [S~II] 6716/6731 emission line ratio maps (shown in Figure 8). The [S~II] 6716/6731 ratio is a direct diagnostic of the electron density of the emitting gas, and  the [S~II] (6716+6731)/H$\alpha$ ratio is a measure of the excitation state of the gas and can be used to estimate the values of the shock velocity that dominate the emission along a given line of sight.  We have constructed the corresponding map of the electron density structure seen in the collimated jet by using the [S~II]~6716/6731 line ratios to solve the 5-level atom problem.  We assumed a uniform electron temperature of $10^4$~K in the emitting region and used the nebular.temden routine in IRAF to calculate the densities.

The [S~II]~6716/6731 and [S~II]/H$\alpha$ line ratio maps show spatial structure along the direction of the outflow (upper panels of Figure 8).   The [S~II]/H$\alpha$ ratio map has several maxima and minima along the jet axis, as well as a general trend of increasing [S~II]/H$\alpha$ ratios with distance from the outflow source (off the panels to the right in Figure 8).  This was first discovered in the longslit spectra presented by Cohen \& Jones (1987).  The E knot shows a ratio value of 5-6, whereas the J knot region has values of 8-10.  Knot I appears to have significant spatial structure in the [S~II]/H$\alpha$ line ratio in the direction perpendicular to the outflow axis.  The relatively large values of the [S~II]/H$\alpha$ ratio imply that the excitation of the jet is rather low (Cohen \& Jones 1987; B\"uhrke, Mundt \& Ray 1988).  However, some of the values measured from our observations are significantly larger than the [S~II]/H$\alpha$ ratios reported by Cohen \& Jones (1987) and B\"uhrke, Mundt \& Ray (1988).   

Perhaps the most striking variation in the line ratios are the curious ``striped'' or ``banded'' structures seen in the [S~II]~6716/6731 map (Figure 8).  The line ratio shows pronounced oscillations, with higher values in the regions to the right of each of the intensity peaks and minima to the left of the successive knots.  Larger line ratio values correspond to lower electron densities in the emitting regions, and smaller values correspond to larger densities.  These observations show convincingly that higher electron densities are found in each of the E through J knots toward the left of the emission peaks.  The locations of higher n$_e$ correspond to the leading edges of the bow shocks, and the minima are found in the trailing sides of the bow shocks, in the ``down-flow'' direction (the exciting source is off the field to the right in Figure 8).  The largest variation in line ratio is seen in knots I and J.  The [S~II] ratios are $\sim$1.20 to the right of the knots, and they decrease to values of $\sim$0.65 and $\sim$0.85 toward the left sides of knots I and J, respectively.  The decrease of the [S~II] ratio from $\sim$1.20 to $\sim$0.65 seen across knot I from right to left implies an abrupt increase of the electron density of the emitting gas from $\sim$ 180 cm$^{-3}$ in the trailing region to 1800 cm$^{-3}$ at the leading side of the emission knot.  In knots E, I and J, the higher electron density regions have approximately symmetric, bow-shaped structures.  The F and G knots both show density maxima that are spatially offset toward alternating sides of the jet axis. 

From the position-velocity cubes of the [S~II] 6716, 6731 \AA\ emission we have generated the electron density channel maps shown in Figure 9.   In each of the channel maps, the electron density also shows the structure of ``bands'' that span the width of the HH 34 jet, as seen in Figure 8 from the integrated electron density.  The $n_e$ channel maps at heliocentric velocities from $-$150 to $-$60 km s$^{-1}$ show very clearly that the high electron density regions correspond to the leading side (left) of each of the knots.  Comparison of the different channel maps presented in Figure 9 shows that the $n_e$ of the low density region increases as the radial velocity becomes less blueshifted (from the low-limit value of $\sim$160 cm$^{-3}$ at $v_r$$\approx$ $-$150 km s$^{-1}$ up to $\sim$350 cm$^{-3}$ at $v_r$$\approx$ $-$60 km s$^{-1}$). The $n_e$ values of the highest density regions remain almost constant at $>$2000 cm$^{-3}$ for radial velocities ranging from $-$120 to $-$60 km s$^{-1}$. 

\subsection{Interpretation}

The P-V diagrams from the observations of B\"uhrke, Mundt \& Ray (1988) showed that the [S~II] emission in the E - K knots have two velocity components.  The brighter emission component was at about $-$85~km~s$^{-1}$ and it remained roughly constant in velocity throughout the extent of the 7 knots.  The second, slower component, was seen in only the G - J knots and it had a radial velocity of $-$53~km~s$^{-1}$ at about one-sixth of the intensity of the fast component.  Heathcote \& Reipurth (1992) showed from their longslit observations that the fast component had a roughly constant velocity of $-$100~km~s$^{-1}$ (referenced to the +27~km~s$^{-1}$ velocity of the nebular material), and that the brightness of the slow velocity component nearly equaled the fast velocity component in the G knot of the HH 34 jet.  Davis et al. (2003) found that the [Fe~II] profiles also exhibited a double-peaked profile shape and were qualitatively similar to the profiles seen in [S~II].   Our observations were of comparable spectral resolution to these optical studies, and we find that the H$\alpha$ feature has a two component profile in many of the knots.  Yet, we find no evidence for a double-peaked profile shape in either the [S~II] 6716 or [S~II] 6731 emission features in the HH 34 jet.  While high spatial resolution images have revealed time variability in the morphology and luminosity of the knots (Reipurth et al. 2002), the accumulated spectral data may now show evidence for temporal variations in the radial velocity kinematics of [S~II] emission in the HH~34 jet.  Variability of emission line profiles has also been seen in the HH~2 outflow (Riera et al. 2005).

The observed variations in the line barycenter radial velocities of the collimated jet (Figure 6) are consistent with the expected velocities for knots that are formed by a succession of ``internal working surfaces''.  The true spatial velocity of the knot material, obtained by deprojecting the barycenter radial velocities, are 10-20~km~s$^{-1}$ larger than the velocities of the inter-knot regions.  This is in surprisingly good agreement with the models of Raga \& Noriega-Crespo (1998).  Their study used analytic internal working surface calculations with the proper motions and spacings of the knots of the HH~34 jet to deduce an expected velocity variability with a half-amplitude of $\sim 10$~km~s$^{-1}$.

The fact that the barycenter velocity decreases outwards from the jet axis to the limb of the jet (see Figure 6) is also expected from models in which the edges of the jet beam have protruding bow shock wings.  In this scenario, the extended wings push the surrounding material to velocities of only a fraction of the jet velocity.   The behavior of the velocity dispersion as a function of position along the jet axis (see Figure 7), with minima just ahead of each of the knots and maxima trailing the flux intensity peaks of the successive knots, is qualitatively consistent with the predictions from bow shock models (Raga \& B\"ohm 1986).  However, the axis-to-limb increase observed for the velocity dispersion of the [S~II] and H$\alpha$ lines (Figure 7) is in principle not compatible with models where the emission comes from bow shocks.  Such models invoke a single, curved shock with the maximum velocity dispersion at the position of the jet axis (e.~g., Hartigan, Raymond \& Hartmann 1987).  The fact that we observe a lower on-axis velocity dispersion could be interpreted as evidence for an increasingly chaotic motion away from the jet axis. 

 Recent high spatial and spectral resolution studies of the emission structure in the inner collimated jets from DG Tau and RW Aur (Bacciotti et al. 2002; Coffey et al. 2004; Woitas et al. 2005) have found evidence for asymmetry in the velocities perpendicular to the outflow axis.  These velocity shifts are  interpreted as evidence for rotation in the jets.  From our observations of HH~34, we find no evidence for shifts in the velocity in the cross-jet direction.  We concede that rotation of the HH~34 outflow could exist below the spectral sensitivity of our observations.  However, for the large distances from the exciting source for the HH~34 collimated jet (thousands of AU), it is more likely that any signature of rotation has long been lost in the turbulent velocity structure inherent to the jet.

Variations in the electron density structure over the E through J knots in the HH~34 complex were first studied by B\"uhrke, Mundt \& Ray (1988).  Their observations, and the longslit spectra presented by Heathcote \& Reipurth (1992), showed that the electron density varies by nearly a factor of 2 over these knots.  Yet, the data presented in these longslit spectroscopic studies did not have the spatial resolution necessary to derive electron density structure within each of the knots themselves.  The [S~II]~6716/6731 line ratios along the HH~34 jet axis that we present show that peaks in the electron density appear in the leading regions of each of the observed knots, and troughs in the density follow each knot.  For the I knot, the variation in electron density from the up to down-flow regions is greater than a factor of five, and the peak density region shows a bow-shaped morphology (Figure 8).  Knots F and G have electron density peaks that are offset from the central axis of the jet, indicating substantial deviations from axisymmetry in this region of the flow (i.e. Reipurth et al. 2002).     These  knots both show density maxima that are spatially offset toward alternating sides of the jet axis and the offsets seem to be roughly consistent with knot positions in the high spatial resolution [S~II] images of Reipurth et al. (2002). Comparison with the HST images of HH~34 of Reipurth et al. (2002) suggests that the electron density maxima coincide approximately with the up-flow sides of the [S~II] emission peaks in the high spatial resolution maps.

Theoretical models of periodic, time variable ejecta from the exciting source are successful at explaining the structure and morphology of the multiple knots and bow shocks of the inner HH 34 outflow (Raga \& Noriega-Crespo 1998; Raga et al. 2002c).  The knots that make up the collimated jet are well described by internal working surfaces that arise when material in the flow plows into slower jet gas.  Curiously, the radial velocity models of Raga \& Cant\'o (1998) predict specifically that there should exist strong electron density jumps of greater than five from up-flow to down-flow regions across internal working surfaces in the knots of the HH 34 jet.  The numerical simulations calculated by Masciadri et al. (2002b) for the HH~111 flow which use a sawtooth ejection variability also predicted density stratifications along the jet axis.  The calculated jumps in electron density across working surface boundaries within an emission knot are precisely what we observe in our spectra.  Our integral field spectroscopic observations of HH 34 have allowed for direct comparison with the existing theoretical models, and have confirmed the predicted existence of electron density jumps within the individual knots in HH jets.

\section{Results: HH~34 Exciting Source}

The six emission knots of the HH 34 A-knot complex are presented in our plots of the exciting source position (Figures 10 - 18).  Figure 10 shows the integrated intensity map constructed by summing the H$\alpha$ (greyscale)  and [S II] 6731 \AA\ (contours) emission lines over their wavelength extent from the data cubes centered on the HH 34 exciting source.  The observed region includes the faint part of the jet extending 5$''$ from the exciting source and the compact optical nebula associated with the object.  The emission from the nebula is bright in H$\alpha$ and shows a V-shaped morphology.  In the H$\alpha$ integrated map we have only detected the innermost knots of the jet (from A6 to A4), while all knots (from A6 to A1) have been detected in the [S II] emission line maps. We have identified the observed knots in Figure 10 following the nomenclature of Reipurth et al. (2002).  As seen in Reipurth et al. (2002), we also find that the peak emission in the exciting source is offset in the H$\alpha$ and [S~II] images.  The highest flux from H$\alpha$ arises from knot A6, and [S~II] emission is most pronounced in the A5 knot.

\subsection{Radial Velocity and Velocity Dispersion}

Figures 11 and 12 present the H$\alpha$ and  [S~II] 6716 \AA\ velocity channel maps of the HH~34 exciting source region.  In the H$\alpha$ channel maps, the compact central emission is detected at heliocentric radial velocities ranging from $-$160 to $+$300~km~s$^{-1}$ (see Figure 11).  Local nebular emission is detected at velocities between $-$10 and $+$30~km~s$^{-1}$ in H$\alpha$, but it is not detected in the [S~II] velocity channel maps.  For both emission species, the innermost part of the outflow knots (A6 to A4) are detected at heliocentric radial velocities ranging from $-$65 to $-$110~km~s$^{-1}$. The H$\alpha$ surface brightness of these knots (A6 to A4) has a maximum emission level in the $-$81~km~s$^{-1}$ velocity map.  The  [S~II] emission from the jet is detected at heliocentric radial velocities spanning from $-$40 to $-$150~km~s$^{-1}$. The surface brightness of the observed knots (knots A6 to A1) increases as $v_r$ becomes more blueshifted, reaches a maximum level at $\sim-$100~km~s$^{-1}$ and then decreases to nearly zero by $v_r\sim-150$. 

We have also computed the H$\alpha$ and [S~II] velocity of the barycenter for the position of the HH~34 exciting source (shown in  Figure 13).  In the upper panel of Figure 13, the H$\alpha$ intensity peak corresponds to the position of knot A6 and has a barycentric velocity of $\sim$ 50~km~s$^{-1}$.  The H$\alpha$ velocities show dramatic variations over the extent of the emission, with changes of $\sim$ 70~km~s$^{-1}$ within a distance of 1$''$ from the exciting source.  Moreover, the velocities change from positive values of $\sim$50~km~s$^{-1}$ on the right side of the outflow source to values of $<-$20~km~s$^{-1}$ to the left.   Conversely, the [S~II] velocity barycenter has small changes along knots A6 to A1.  The slowest velocity is detected at knot A6 and has a value of $-$80~km~s$^{-1}$.  At larger distances the velocity becomes more blueshifted, reaching a value of $-$110~km~s$^{-1}$ between knots A5 and A4.  At the positions of knots A3, A2 and A1 the velocity along the central axis remains almost constant at $\sim-$95~km~s$^{-1}$.  Perpendicular to the jet axis the [S~II] barycenter velocities show a decrease toward less blueshifted values outwards from the jet (as also observed in the collimated region of the HH 34 jet; see Section 3.1).  The difference between the axial velocities and the velocities at the outer edge of the detected jet beam are of $\Delta v_{r,\perp}\approx$ 5 to 10~km~s$^{-1}$ (depending on the spatial position). 

Derivation of the velocity barycenter provides an important means to characterize the overall velocity structure in the exciting source, but this method of analysis is not ideal for distinguishing between multiple emission components.  Three-dimensional isosurface contour plots of the x,y and velocity datacube provide a powerful method of viewing the HH~34 exciting source.  The isosurface contour plots for the H$\alpha$ and [S~II] 6731 features are presented in Figure 14.  These plots were made with the ``slicer3'' 3-D visualization GUI tool inherent to the IDL programming language using contour levels of $2\times10^{-17}$erg cm$^{-2}$ s$^{-1}$ \AA$^{-1}$ (spectral pixel)$^{-1}$ and $4\times10^{-17}$erg cm$^{-2}$ s$^{-1}$ \AA$^{-1}$ (spectral pixel)$^{-1}$  for the H$\alpha$ and [S~II] 6731 features, respectively.  In the H$\alpha$ plot, the component that extends from $\sim-$110~km~s$^{-1}$ to $\sim+$160~km~s$^{-1}$ roughly corresponds to the emission at the position of the A6 knot.  The successive knots of the outflow extend towards the left in the plot (as in Figure 10).  We find from the lower panel in Figure 14 that the isosurface contour plots of [S~II] 6731 are very similar to the velocity barycenter diagram presented in Figure 13;  the largest velocity is seen at knot A6, the most negative value arises from knots A5 and A4, and the A3 through A1 knots show essentially constant velocities.  

The H$\alpha$ isosurface contour plot shown in the upper panel of Figure 14 exhibits a much more complicated velocity structure with two or three velocity components.  The local nebular emission, as mentioned previously, is a dominant component in the $v_r$$\sim$$-$10 to $+$30~km~s$^{-1}$ velocity region.  At the position closest to the exciting source (A6), a broad and pronounced emission component ranges from velocities of $-$110~km~s$^{-1}$ to $\sim+$160~km~s$^{-1}$.  At the positions of the A4 and A5 knots, an isosurface velocity component with a roughly constant $v_r$$\sim$-100~km~s$^{-1}$ is detected.  This blueshifted emission dominates the flux at the position of the A5 knot and is the reason for the pronounced shift in barycenter velocity from knot A6 to A5 (upper panel of Figure 13).  No corresponding redshifted counter-jet is seen in HH~34 in the isosurface contour plot.

Figure 15 shows the velocity dispersion maps constructed around the position of the HH~34 source. The velocity dispersion seen in H$\alpha$ is consistent with the complex velocity character seen in the emission from the isosurface map in Figure 14.  The [S~II] velocity dispersion map shows small variations along the central axis, and has a value of $\sim$ 46~km~s$^{-1}$ at knot A6.  The  [S~II] velocity dispersion reaches a minimum value of $\sim$ 30~km~s$^{-1}$ between knots A5 and A4, but at larger distances from the central source it increases again to values of $\approx$ 45~km~s$^{-1}$.  The dispersion appears to have slightly larger values away from the central outflow axis at knots A1 and A2, as was seen in section 3.1 for the collimated outflow.

\subsection{Emission Line Ratios}

The integrated intensity maps of the [S~II] 6716, 6731 \AA\ and H$\alpha$ emission lines were obtained by summing each line over its wavelength extent in the data cube. From these intensity maps we have constructed the [S~II] (6716+6731)/H$\alpha$ and  [S~II] 6716/6731 emission line ratio maps shown in Figure 16, and Figure 17 presents the corresponding electron density map.  The [S~II] 6716/6731 emission line ratio and the electron density maps illustrate the strong increase of electron density towards the source, as already pointed out by Cohen \& Jones (1987), B\"uhrke, Mundt \& Ray (1988), and Heathcote \& Reipurth (1992).  The innermost knots (A6-A4) are emerging from the high density material surrounding the central source.  At these knots we measure electron densities from $>$10$^{4}$~cm~$^{-3}$ to 7500~cm~$^{-3}$ (at knot A5).  The [S~II] 6716/6731 line ratio increases and therefore, the electron density decreases, with increasing distance from the outflow source.  In the region sampled by our IFU spectroscopy of the exciting source, the electron density has a value of $\sim$ 1200~cm~$^{-3}$ at the position of knot A1, but it decreases to 500-600~cm~$^{-3}$ to the left of this knot. 

The integrated [S~II]/H$\alpha$ line ratio map (Figure 16) shows a strong variation along the central axis of the jet, with a general trend of increasing  [S~II]/H$\alpha$ with distance from the outflow source.  The [S~II]/H$\alpha$ ratio has a value of 1.0 at knot A6.  This is not surprising because although the peak flux of the [S~II] emission is greater at the position of the exciting source, the total H$\alpha$ emission is comparable because the emission feature is very broad. Moving down the axis of the jet in the [S~II]/H$\alpha$ map (away from the source), we see a rapid increase of this line ratio as the contribution from the central H$\alpha$ flux decreases.  The ratio reaches a value of 4.6 at knot A1.  The H$\alpha$ emission from knots A3 to A1 is considerably less than that for [S~II], but it is detected at a weak level of significance.  We do find evidence for a change in the  [S~II]/H$\alpha$ line ratio in the cross-jet direction (at the position of knots A1 and A2).   Away from the central source, the [S~II]/H$\alpha$ line ratio consistently shows a decrease outwards from the jet axis.  

Figure 18 shows the electron density channel maps obtained from the [S II] 6716/6731 line ratio velocity maps. Each of these maps shows the rapid decrease of the electron density with distance from the outflow source.  The greatest electron density of $\approx$ 2$\times$10$^{4}$~cm$^{-3}$ is found at the position of knot A5 in the  v$_r =$ $-$74 to $-$104~km~s$^{-1}$ velocity maps.  Comparison of the different channel maps shown in Figure 18 shows consistently the increase of the electron density with heliocentric radial velocities increasing from $-$134 to $-$74~km~s$^{-1}$.  The electron density then decreases at all positions for less blueshifted radial velocities.

\subsection{Interpretation}

The ``A Knot'' of the HH~34 jet was first identified and named in the studies of Reipurth et al. (1986) and Cohen \& Jones (1987).  It was found by Ray et al. (1996) to be comprised of a conglomeration of several smaller knots when viewed at high spatial resolution.   For this study, we have identified the A6 knot to be where the H$\alpha$ emission is most intense (Figure 10).  However, Reipurth et al. (2000) and Rodr\'\i guez \& Reipurth (1996) showed that the infrared and radio positions of the exciting source are actually not coincident with the A6 knot detected in the optical.  The exciting source is known to be highly obscured, the optical emission at knot A6 likely arises from either scattered light off the wall of the inner outflow cavity or emission from the inner-most knot of the outflow.  Hence, in our observations we do not actually see the star, which is revealed only at longer wavelengths.

Figures 3 and 14 show that the velocity structure near the exciting source position has a very strong, broad  H$\alpha$ emission component that is roughly symmetric in velocity and not present in [S~II].  Such emission features have been seen in many embedded young stars and arise from the magnetospheric accretion emission of material onto the central star.  The isosurface map presented in Figure 14 shows three distinct velocity components in the H$\alpha$ emission.  The broad emission component is seen at the source position and extends from $\sim-$110 to +160 km/s.  The emission that is spatially extended and centered at a velocity of $\sim$10 km/s is imperfectly subtracted, residual H$\alpha$ arising from the ambient nebulosity in the environment of the HH 34 jet.  The third emission component evident in the H$\alpha$ isosurface plot arises from the blueshifted jet itself; it extends left-ward from the central source in Figure 14 at velocity of $\sim-$100 km/s.  The blueshifted jet of HH 34 was also studied by Takami et al. (2006) and was shown to best support disk-wind models for the outflow.  The three-component velocity structure can not be easily discerned in the velocity barycenter or dispersion plots (Figures 13 and 15).

Unlike for the collimated jet, [S~II] 6716/6731 line ratios and corresponding electron density maps in the inner A knot region of the HH 34 outflow do not show a ``striped'' structure.  The minimum n$_e$ seen in this region is $\sim$500-600cm$^{-3}$ to the left of the A1 knot, and the electron density increases to $>$10$^4$cm$^{-3}$ toward the inner outflow region of the HH~34 jet.  The A4 to A6 knots are within less than $\sim$2$''$ from the central star ($<$900 AU at the distance of HH 34).  The ambient high energy flux and density of parent cloud material are both probably greater near the exciting source.  Thus, the ionization fraction and the gas density are likely greater.  The resulting higher n$_e$ values probably cause the loss of the ``striped'' structure from fluctuations seen in the collimated jet.  Moreover, the electron density has a peak value that is spatially offset from the position of the embedded source; the peak is coincident with the area between the A4 and A5 knots (Figure 17).  X-ray emission from fast shocks, recently identified in the HH~154 and HH~2 outflows, could give rise to high energy ionizing flux in the inner jet region of HH~34 (Pravdo et al. 2001, Favata et al. 2002, Bally, Feigelson \& Reipurth 2003, Raga, Noriega-Crespo \& Vel\'azquez 2002b).  It seems conceivable that shock excited high energy emission in the outflow could ionize the gas and cause an increase in the electron density spatially offset from the exciting source.  To date, no X-ray emission has been detected toward the inner HH~34 jet.

\section{Summary}

In this paper, we have presented the first optical integral field maps of the H$\alpha$ and [S~II] emission associated with the HH 34 exciting source and a $\sim$10$''$ region of the collimated jet.  These high spatial and spectral resolution data show:

1) Velocity and velocity dispersion peaks and troughs exist along the axis of the collimated jet, even in the inner regions of the flow.  The velocity peaks are associated with the brighter emission regions in the jet, whereas the minima in dispersion are typically on the knots.
 
2) In the cross-jet direction, perpendicular to the outflow axis, the velocity decreases outward, but the dispersion increases.

3) The spatially resolved structure in the [S~II] 6716/6731 line ratios along the collimated jet show that high electron density regions are at the front of each emission knot, and low density regions trail the emission peaks.  These data confirm models that predicted electron density jumps across the working surfaces in collimated jets viewed edge-on (Masciadri et al. 2002b; Raga \& Canto 1998).

4) The electron density in the inner regions toward the outflow source increases significantly.  The peak in the electron density is n$_e>2\times10^4$/cm$^3$ and is spatially offset by more than 1$''$ from the position of the exciting star.

5) The data presented here provide a new spatially resolved view of the kinematics and electron density structure of the HH~34 jet.  We find that the kinematic results are largely consistent with the models of HH~34 already presented by Raga \& Noriega-Crespo (1998), Cabrit \& Raga (2000), Masciadri et al. (2002a) and Raga et al. (2002c).  In a future paper, we will carry out a detailed comparison between the data discussed above and predictions from variable jet models.

\acknowledgments

We are grateful to our referee, Hongchi Wang, for reading our manuscript carefully and providing suggestions for its improvement.  We also thank Inger J\o rgensen for her assistance setting up the observations for our project and the Gemini GMOS queue observers for taking data for the program.  The data presented for this project were acquired in Gemini queue observing mode through University of Hawaii and Director's Discretionary time under the Gemini program ID GN-2003A-Q-16.  A. Riera is supported by the MEC AYA2005-05823-C03 grant.  This study was supported by the Gemini Observatory, which is operated by the Association of Universities for Research in Astronomy, Inc., under a cooperative agreement with the NSF on behalf of the Gemini partnership: the National Science Foundation (United States), the Particle Physics and Astronomy Research Council (United Kingdom), the National Research Council (Canada), CONICYT (Chile), the Australian Research Council (Australia), CNPq (Brazil), and CONICET (Argentina).

\clearpage

\clearpage


\begin{figure}
\epsscale{1.0}
\plotone{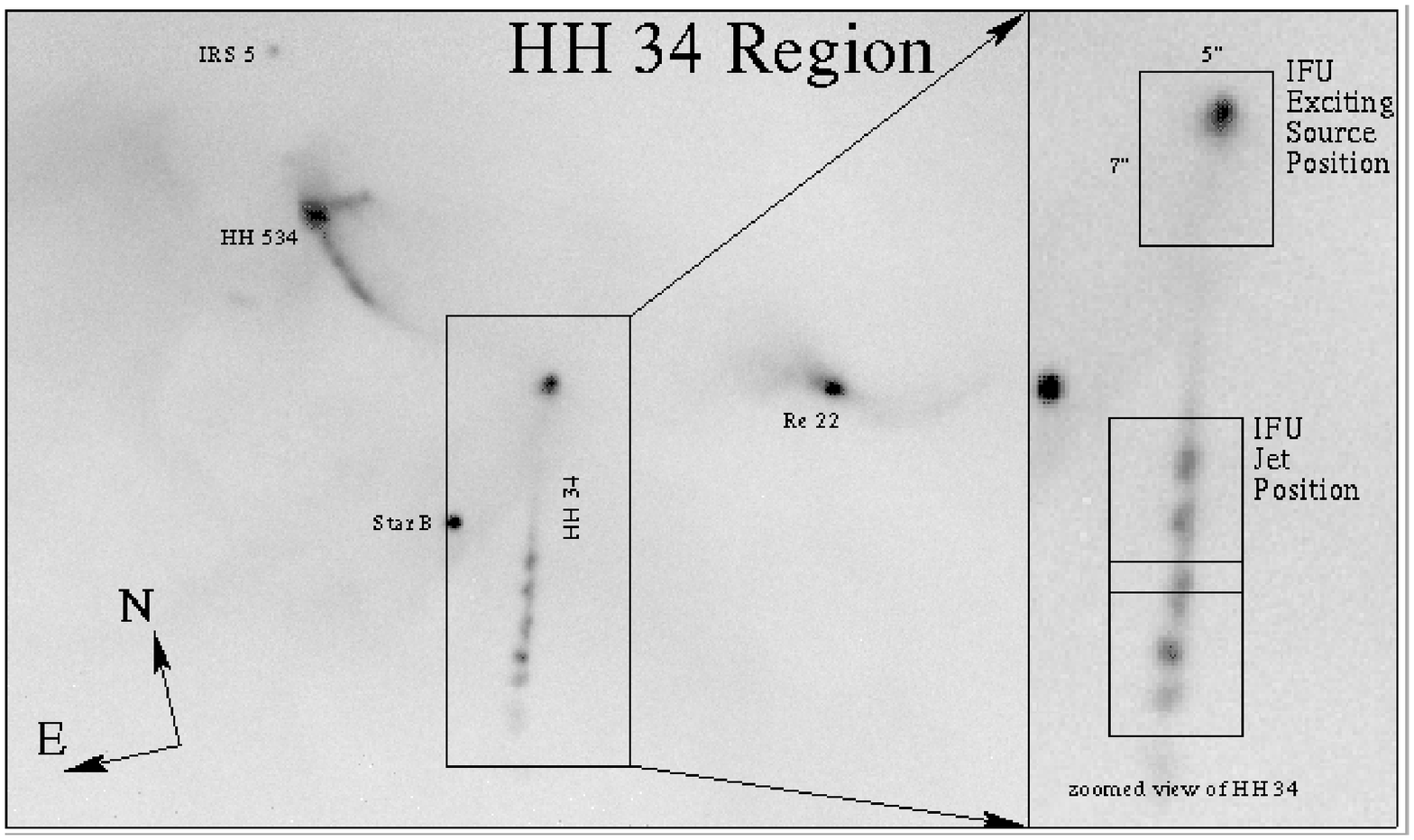}
\caption{An image of the HH 34 region, acquired for setup of the IFU observations.  This image is a median of six 300s exposures obtained through the H$\alpha$ filter in the GMOS instrument. The exciting source of the HH 34 outflow and the collimated jet are identified, as are several other stars in the vicinity.  The orientation of the field is presented in the lower left, and the size of the image is $\sim$66$''$ $\times$ 52$''$.  The rigth panel shows a zoomed view of the HH 34 exciting source and collimated jet.  Overplotted in this image are the outlines of the three rectangular 7$''\times$5$''$ IFU positions used for this study, one position was on the exciting source and two were on the collimated jet.\label{fig1}}
\end{figure}

\begin{figure}
\epsscale{1.1}
\plotone{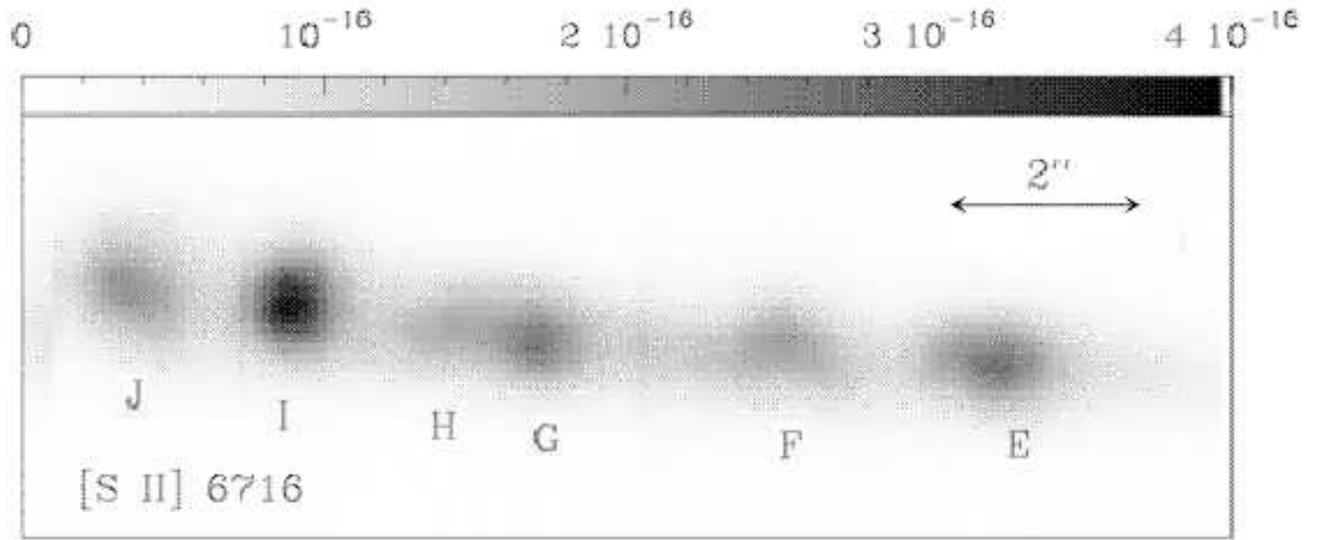}
\caption{[S II] 6716 \AA\  intensity map of the collimated jet of HH 34 constructed by summing the flux over this emission line in the IFU datacube.  The observed knots are identified following the nomenclature of B\"uhrke, Mundt \& Ray (1988).
\label{fig2}}
\end{figure}

\clearpage

\begin{figure}
\epsscale{0.7}
\plotone{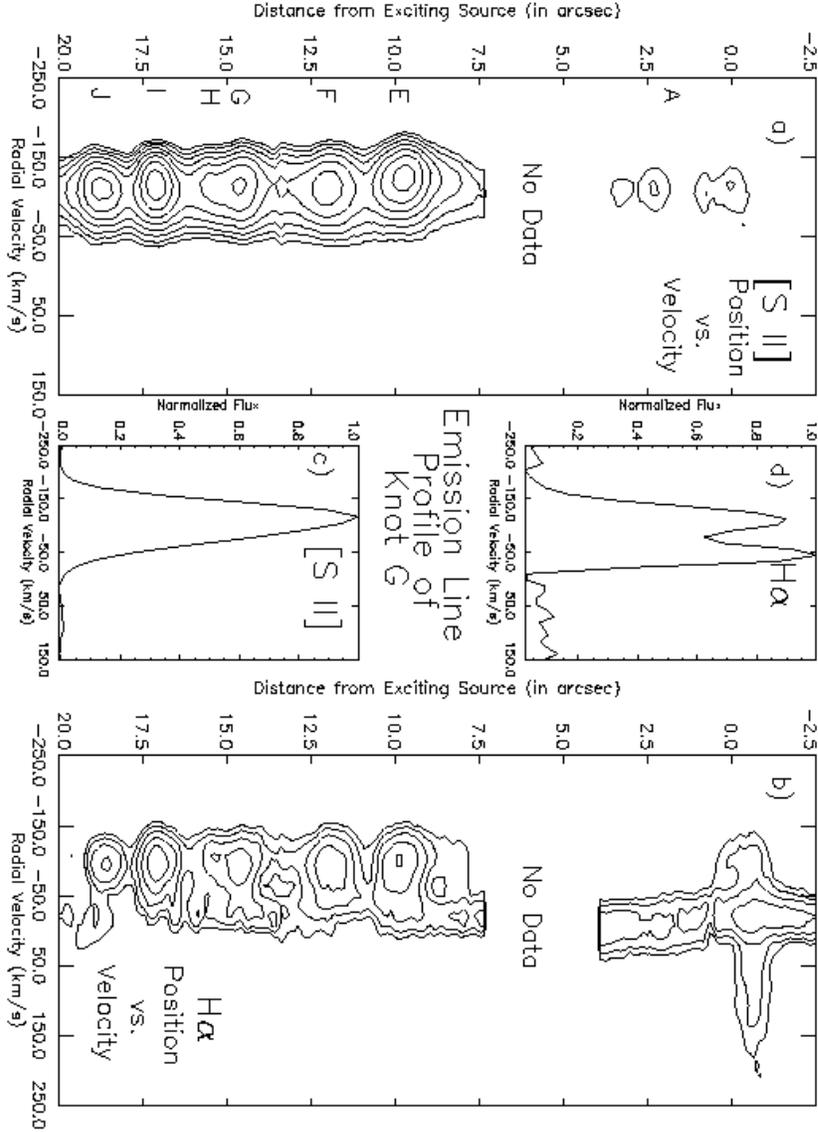}
\caption{The position-velocity diagrams for the [S~II] emission (panel a, to the left), and H$\alpha$ emission (panel b, to the right) of the HH~34 collimated jet and exciting source.  The lowest contour values are $3.0\times10^{-18}$erg cm$^{-2}$ s$^{-1}$ \AA$^{-1}$ (spectral pixel)$^{-1}$ and $1.5\times10^{-18}$erg cm$^{-2}$ s$^{-1}$ \AA$^{-1}$ (spectral pixel)$^{-1}$ for the [S~II] and H$\alpha$ diagrams, respectively.  The contour levels increase in intervals of 2$^{1/2}$.  Plotted in the center in panels c and d are the profiles of the emission line features that were derived from summing the flux at all spatial positions over the G knot.
\label{fig3}}
\end{figure}

\clearpage

\begin{figure}
\epsscale{1.1}
\plotone{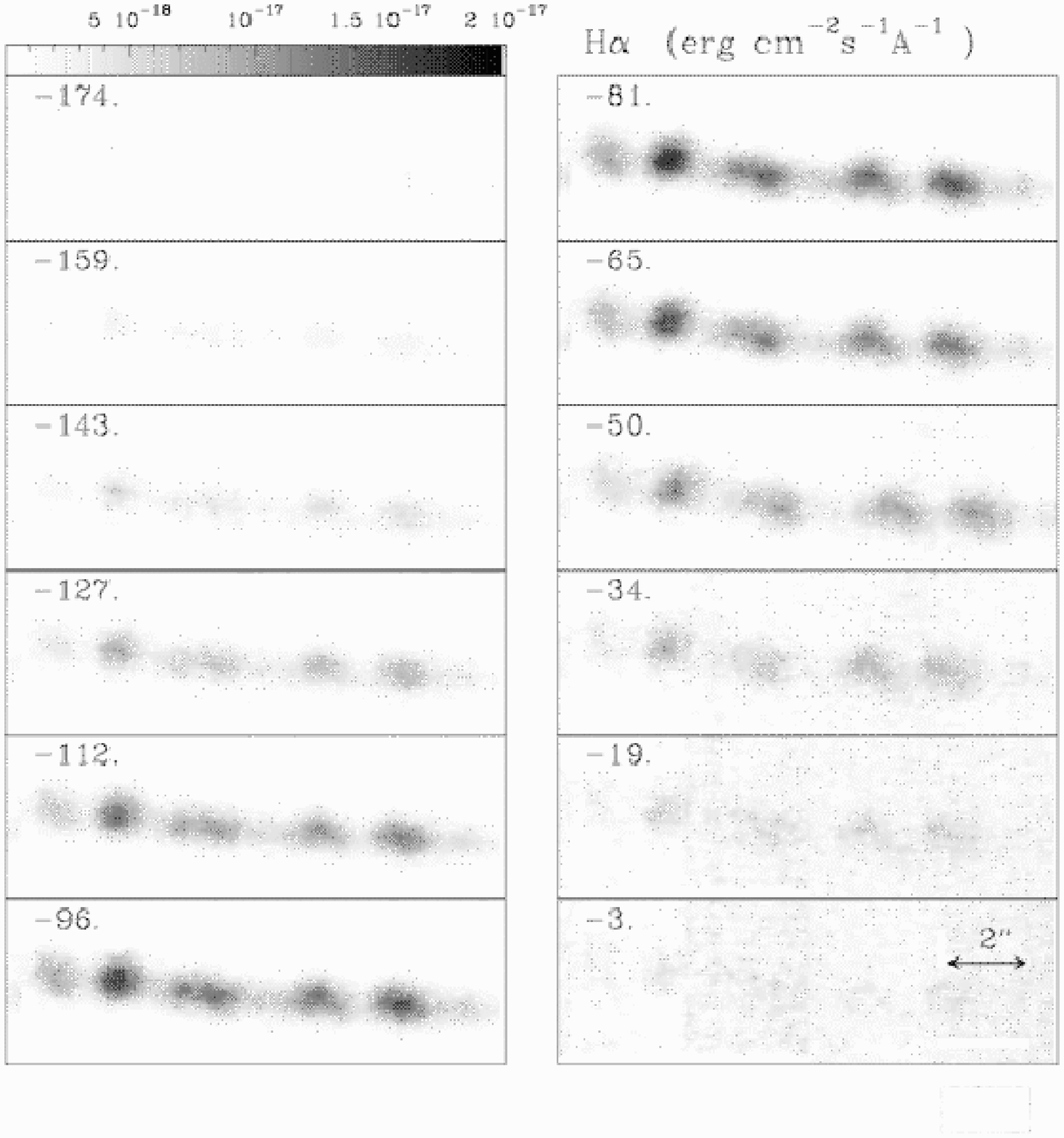}
\caption{H$\alpha$ radial velocity channel maps derived from the IFU spectral datacubes of the collimated HH 34 jet. 
The maps are 
shown with a linear scale given by the bar (in erg cm$^{-2}$ s$^{-1}$ \AA$^{-1}$ pixel$^{-1}$). 
\label{fig4}}
\end{figure}
\clearpage

\begin{figure}
\epsscale{1.1}
\plotone{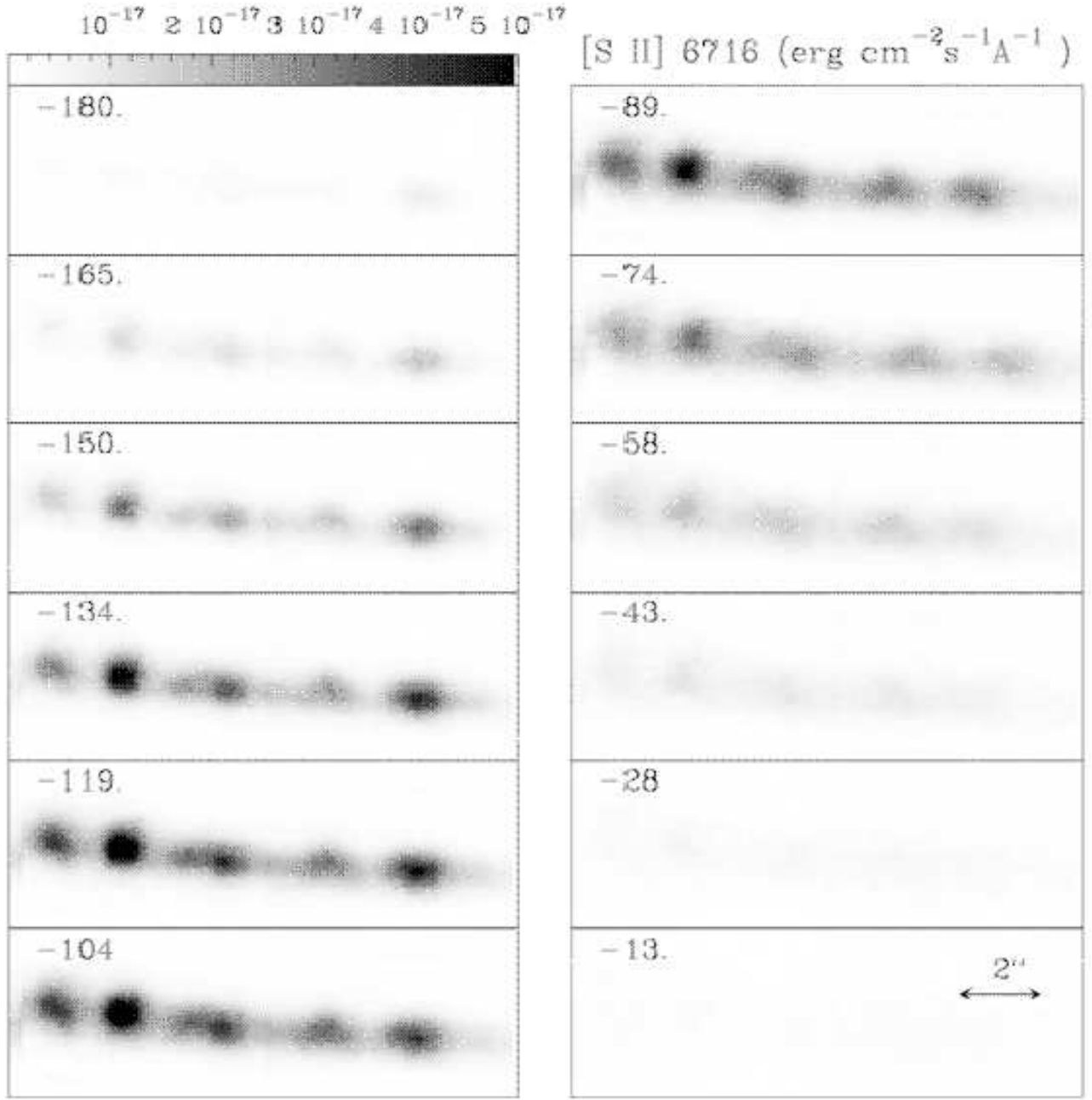}
\caption{[S II] 6716 \AA\ velocity channel maps  derived from the IFU spectral datacubes of the collimated HH 34 jet.
The maps are 
shown with a linear scale given by the bar (in erg cm$^{-2}$ s$^{-1}$ \AA$^{-1}$ pixel$^{-1}$). 
\label{fig5}}
\end{figure}
\clearpage

\begin{figure}
\epsscale{1.1}
\plotone{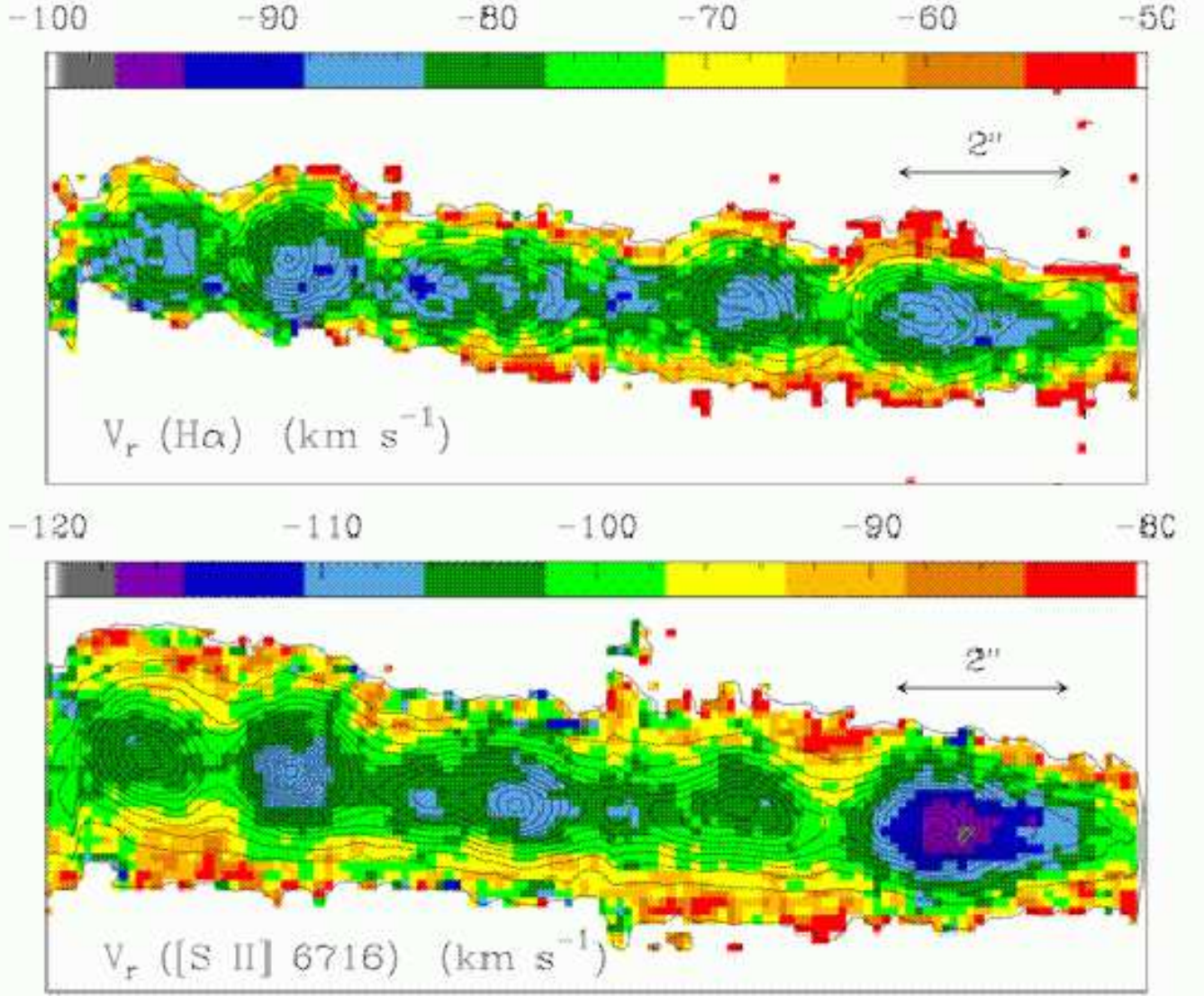}
\caption{H$\alpha$ (upper panel) and [S II] 6716 \AA\ (bottom panel) barycenter radial velocity map obtained from the position-velocity cubes of the collimated HH 34 jet.  The color scales are linear in km s$^{-1}$ and are represented by the bars.  The linear contours correspond to the H$\alpha$ and [S II] 6716 \AA\ integrated flux images. The H$\alpha$ contours span from 10$^{-17}$ to 1.5$\times$10$^{-16}$ erg cm$^{-2}$ s$^{-1}$ \AA$^{-1}$ pixel$^{-1}$ with steps of 10$^{-17}$ erg cm$^{-2}$ s$^{-1}$ \AA$^{-1}$ pixel$^{-1}$. The [S II] 6716 \AA\ contours range from 10$^{-17}$ to 4.0$\times$10$^{-16}$ erg cm$^{-2}$ s$^{-1}$ \AA$^{-1}$ pixel$^{-1}$ with steps of 2.0$\times$10$^{-17}$  erg cm$^{-2}$ s$^{-1}$ \AA$^{-1}$ pixel$^{-1}$. 
\label{fig6}}
\end{figure}
\clearpage

\begin{figure}
\epsscale{1.1}
\plotone{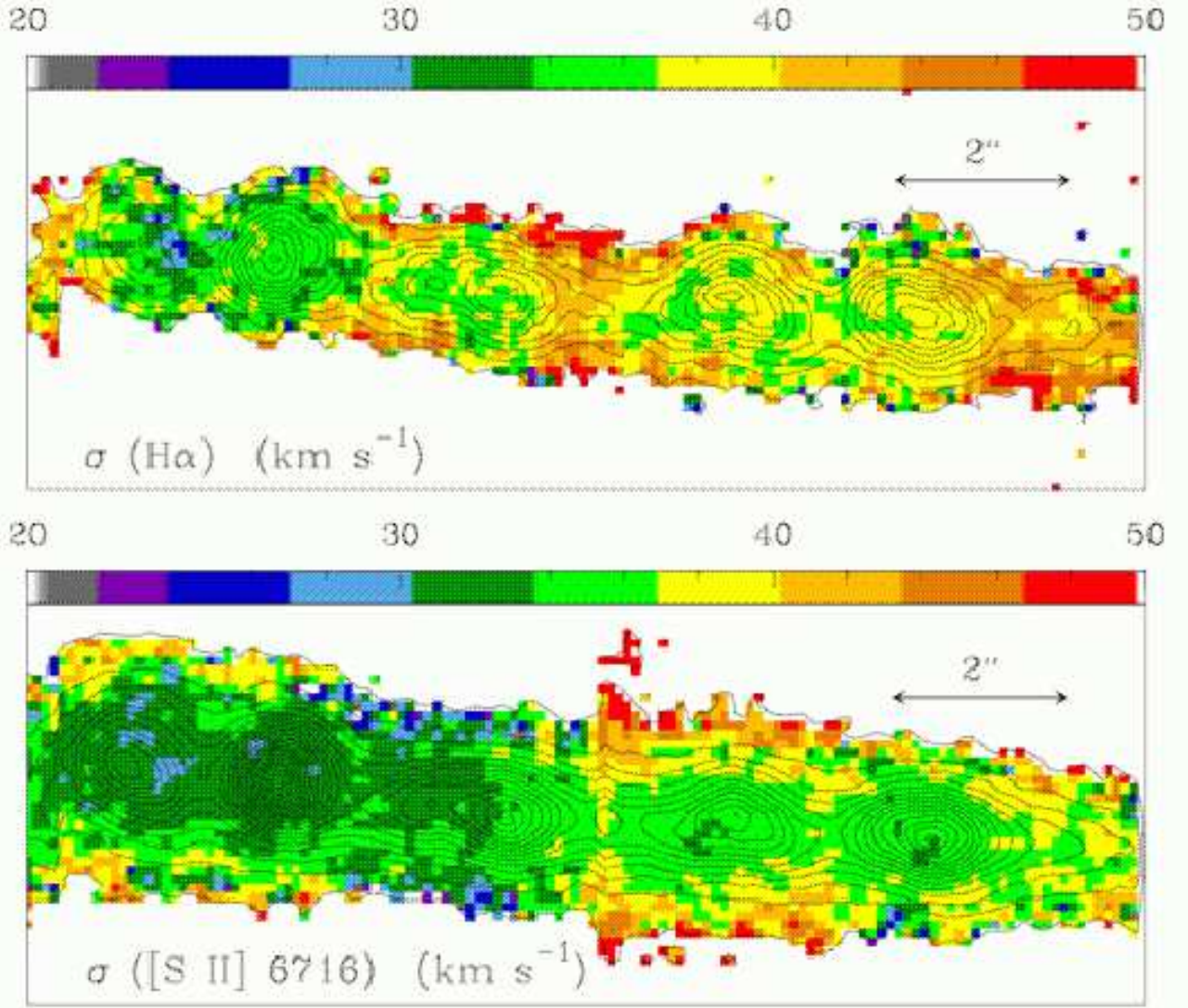}
\caption{H$\alpha$ (upper panel) and [S II] 6716 \AA\ (bottom panel) line dispersion maps obtained from the position-velocity cubes corresponding to the collimated HH 34 jet. The scales are linear in km s$^{-1}$ and are given by the bars.
\label{fig7}}
\end{figure}
\clearpage

\begin{figure}
\epsscale{1.1}
\plotone{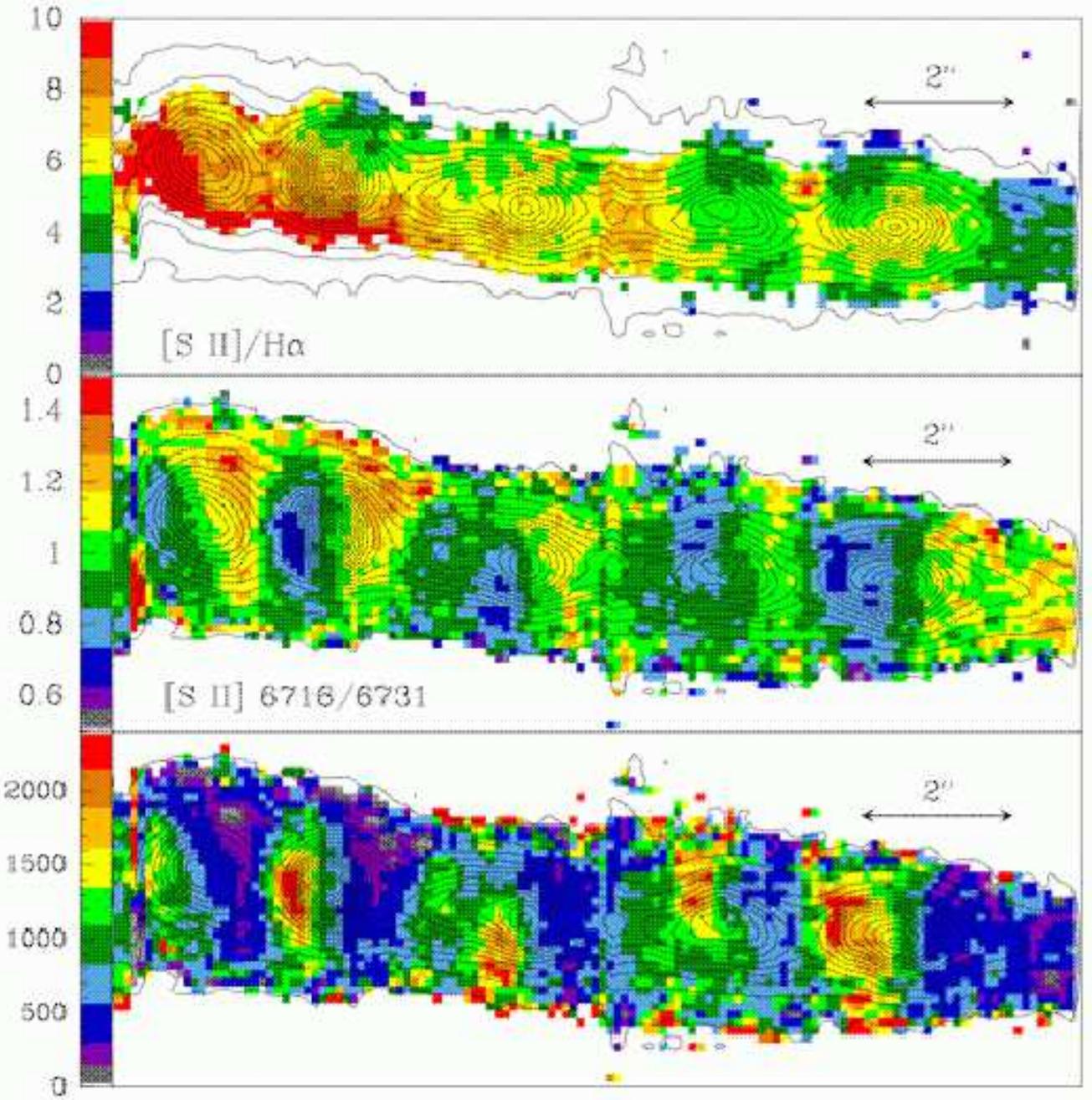}
\caption{[S II] (6716+6731)/H$\alpha$ (upper panel) and [S II] 6716/6731 (central panel) ratio maps obtained by summing the [S II] 6716, 6731 \AA\ lines over its wavelength extent in the data cube. The spatial region corresponds to the collimated HH 34 jet. The scales are linear and are shown by the bars.  Bottom panel: the electron density (n$_e$) map obtained from the [S II] 6716/6731 ratio map. The spatial region corresponds to the collimated HH 34 jet. The scale is in cm$^{-3}$ and is shown by the bar. 
\label{fig8}}
\end{figure}
\clearpage

\begin{figure}
\epsscale{1.1}
\plotone{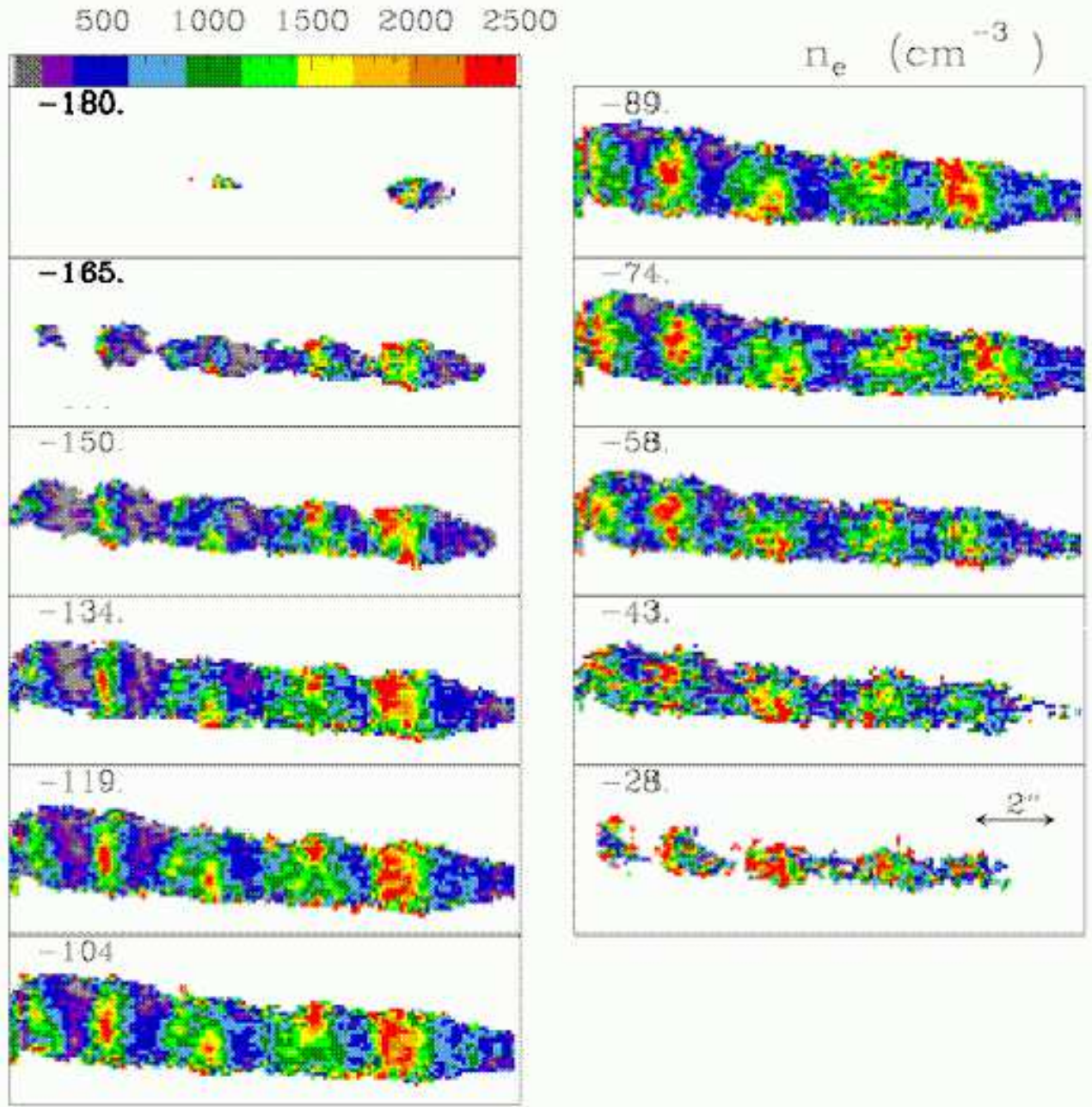}
\caption{The electron density (n$_e$) map derived from the [S II] 6716/6731 ratio maps  at the heliocentric radial velocity shown in each panel.   
The scale in cm$^{-3}$ is shown by the bar.
\label{fig9}}
\end{figure}
\clearpage

\begin{figure}
\epsscale{1.1}
\plotone{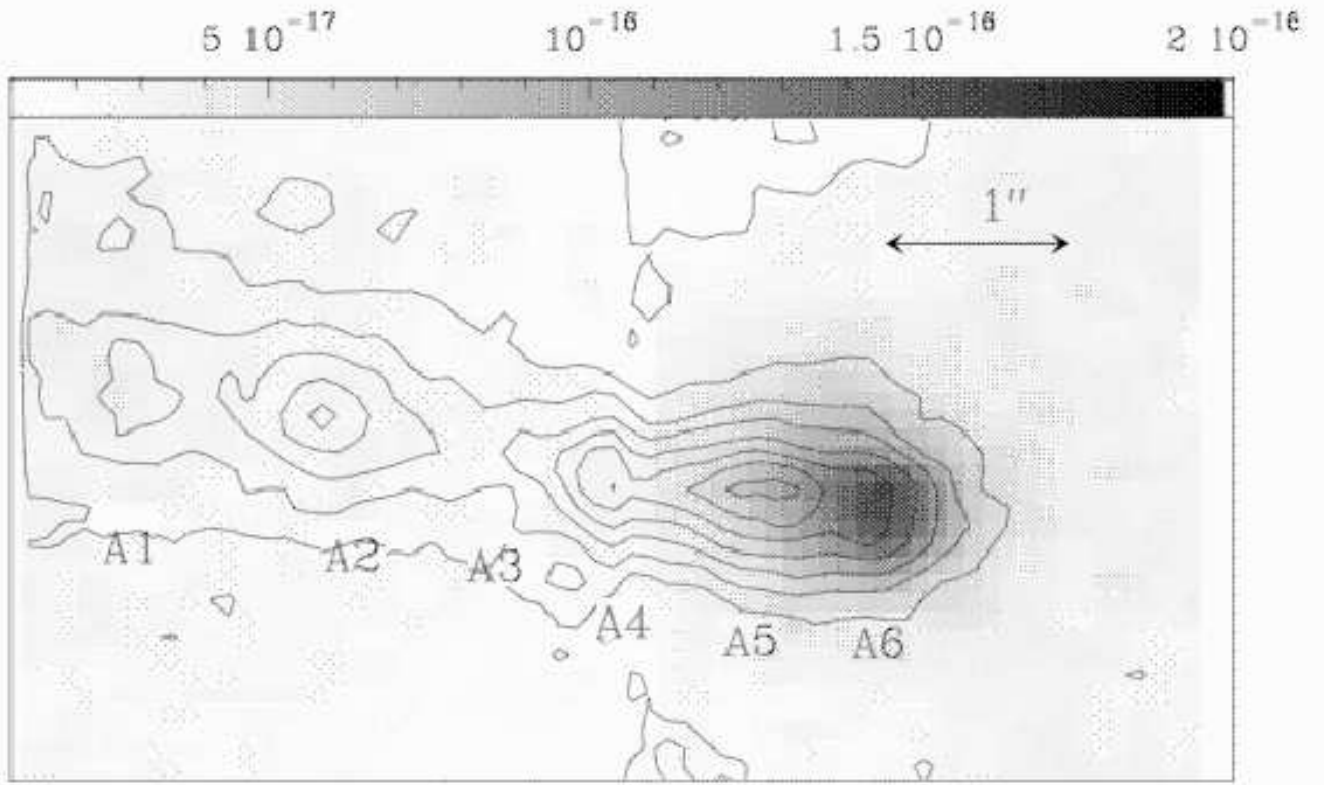}
\caption{H$\alpha$  intensity map of the HH 34 exciting source. The linear contours correspond to the [S II] 6731 \AA\ intensity map. The observed knots are identified following the nomenclature of Reipurth et al. (2002).
\label{fig10}}
\end{figure}
\clearpage

\begin{figure}
\epsscale{1.1}
\plotone{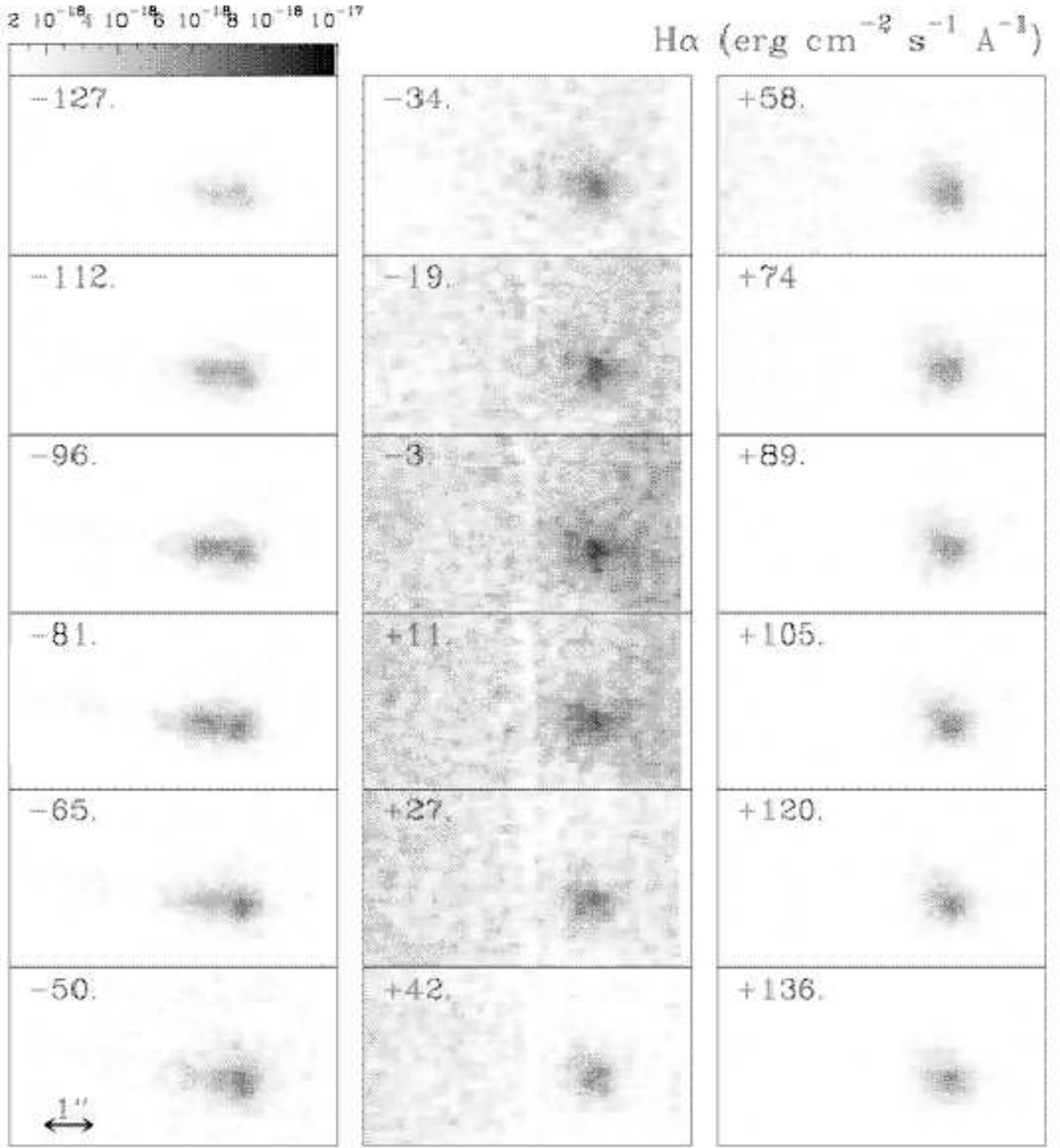}
\caption{H$\alpha$ velocity channel maps obtained with GMOS IFU spectrograph on the HH 34 exciting source region. 
The maps are shown with a linear scale given by the bar (in erg cm$^{-2}$ s$^{-1}$ \AA$^{-1}$ pixel$^{-1}$).
\label{fig11}}
\end{figure}
\clearpage

\begin{figure}
\epsscale{1.1}
\plotone{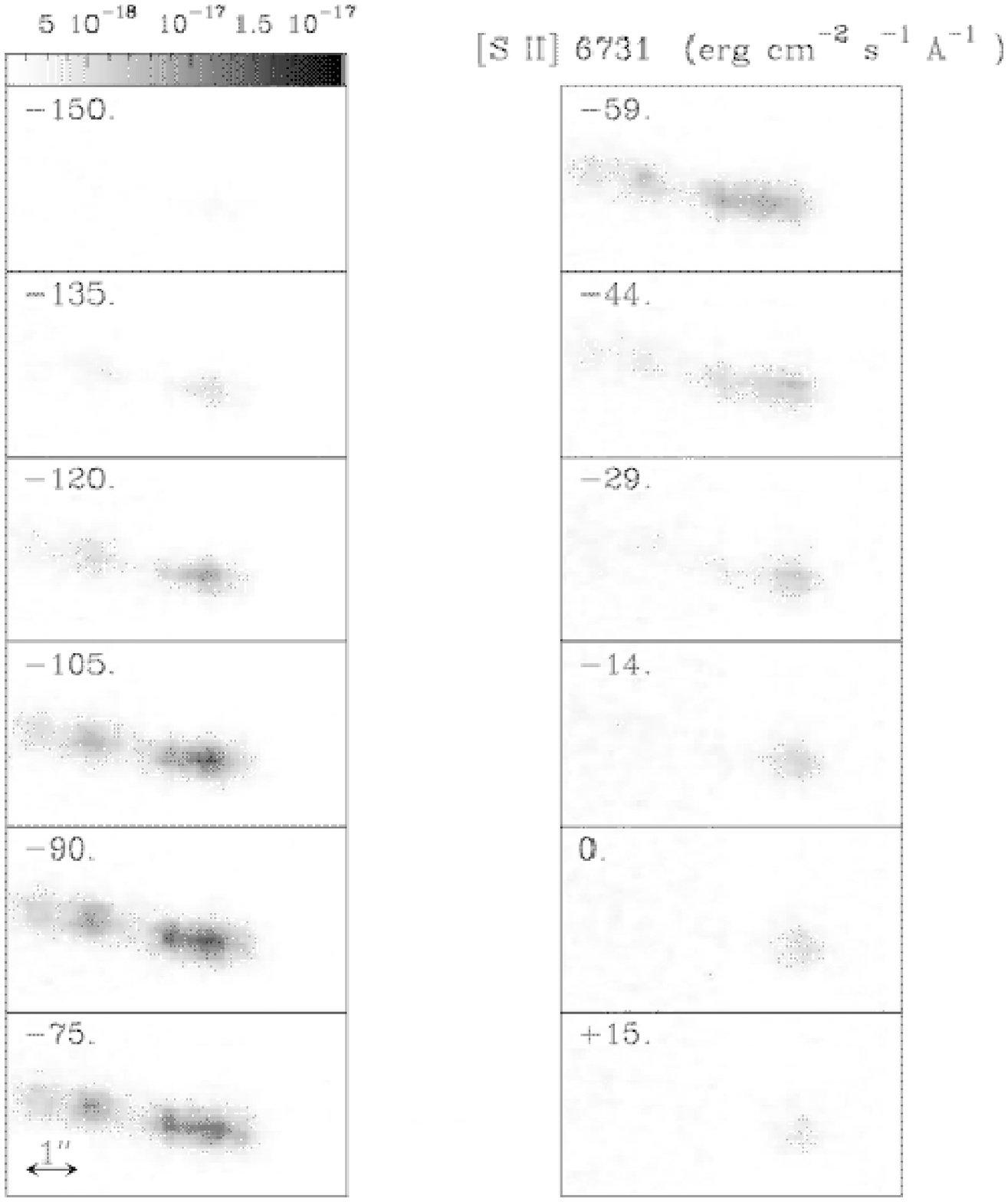}
\caption{[S II] 6731 \AA\ velocity channel maps obtained with GMOS IFU on the HH 34 exciting source region. 
The maps are shown with a linear scale given by the bar (in erg cm$^{-2}$ s$^{-1}$ \AA$^{-1}$ pixel$^{-1}$).
\label{fig12}}
\end{figure}
\clearpage

\begin{figure}
\epsscale{1.1}
\plotone{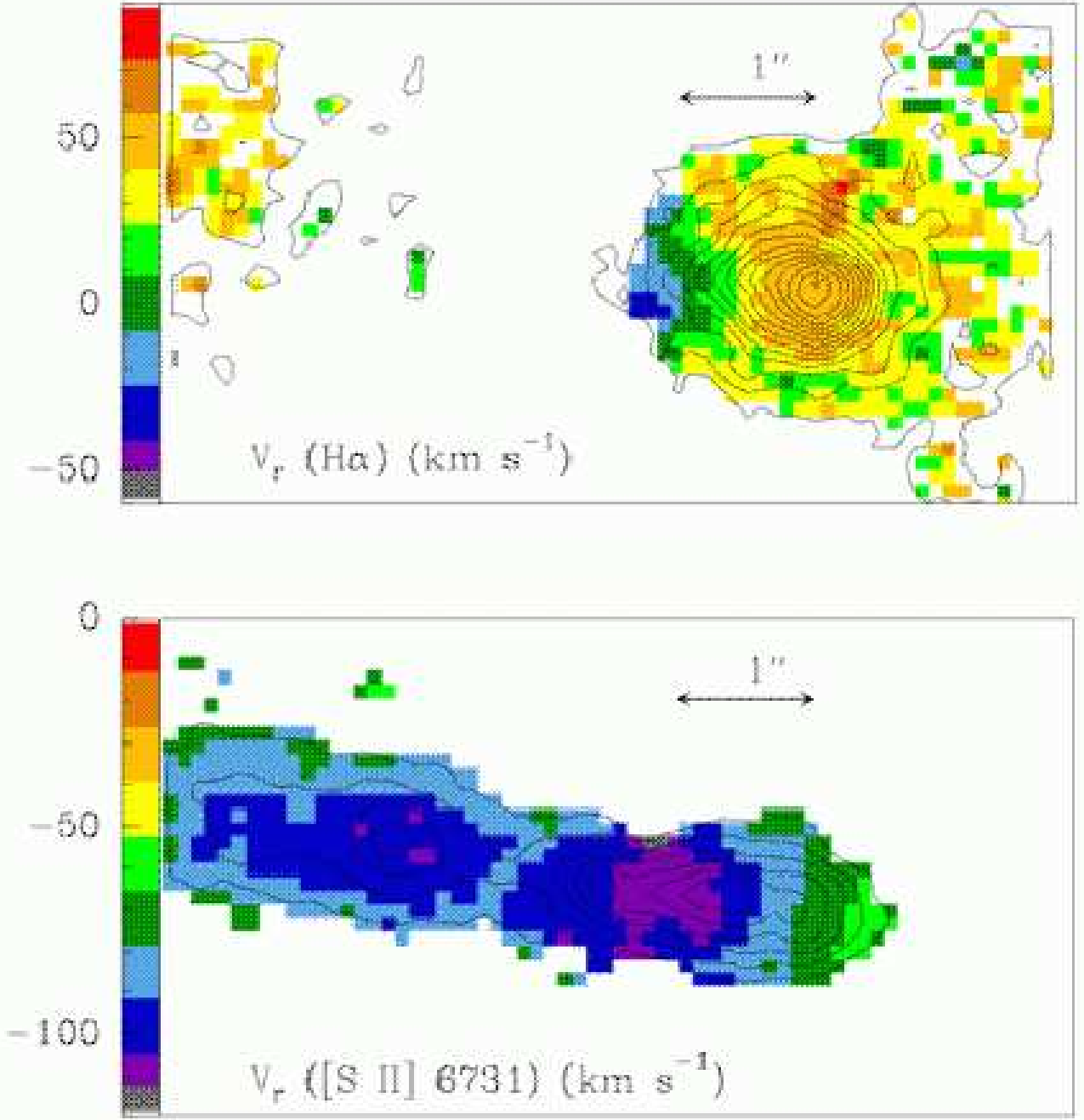}
\caption{H$\alpha$ (upper panel) and [S II] 6731 \AA\ (bottom panel)  barycenter radial velocity map obtained from the position-velocity cubes in the vicinity of the HH 34 exciting source.  The color scales are linear in km s$^{-1}$ and are given by the bars.  The linear contours correspond to H$\alpha$ (top pannel) and [S II] 6731 \AA\ (bottom pannel) intensity maps. The H$\alpha$ contours go from 3$\times$10$^{-17}$ to 2$\times$10$^{-16}$ erg cm$^{-2}$ s$^{-1}$ \AA$^{-1}$ pixel$^{-1}$.  The [S II] contours go from 2$\times$10$^{-17}$ to 1$\times$10$^{-16}$ erg cm$^{-2}$ s$^{-1}$ \AA$^{-1}$ pixel$^{-1}$. 
\label{fig13}}
\end{figure}
\clearpage

\begin{figure}
\epsscale{1.0}
\plotone{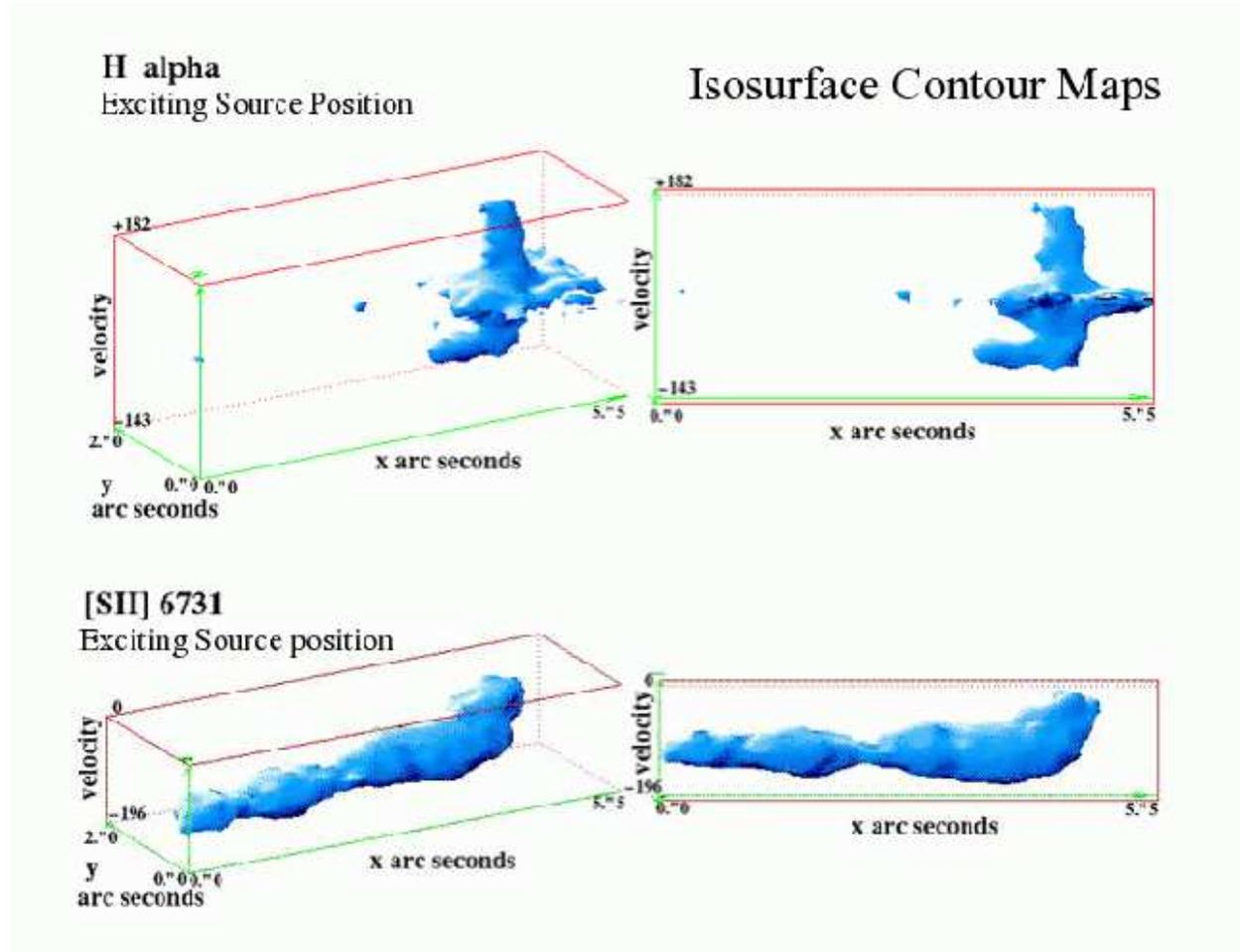}
\caption{The three dimensional isosurface contour plots for H$\alpha$ and [S~II] 6731 \AA\ emission from the position of the exciting source of the HH~34 outflow.  The isosurface in the plot of x position on the sky, y position on the sky, and velocity for  [S~II] is at a flux level of 4$\times$10$^{-17}$ erg cm$^{-2}$ s$^{-1}$ \AA$^{-1}$ pixel$^{-1}$, and for H$\alpha$ the surface is at a flux of 2$\times$10$^{-17}$ erg cm$^{-2}$ s$^{-1}$ \AA$^{-1}$ pixel$^{-1}$.  The spatial axes are in arcseconds, and the velocity axis is in units of km s$^{-1}$. \label{fig14}}
\end{figure}

\begin{figure}
\epsscale{1.1}
\plotone{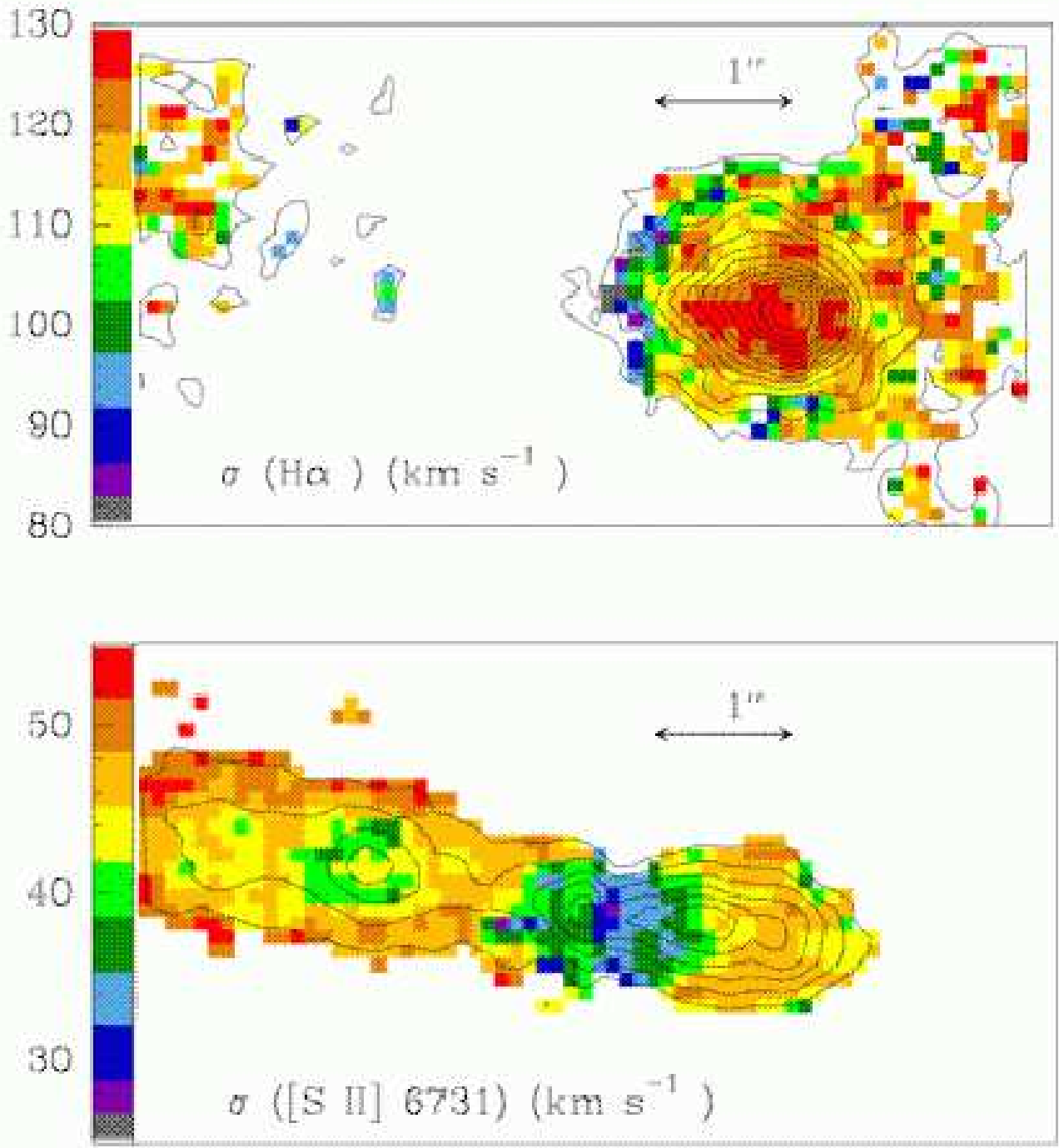}
\caption{H$\alpha$ (upper panel) and [S II] 6731 \AA\ (bottom panel) 
 velocity dispersion map obtained from the position-velocity cubes 
in the vicinity of the HH 34 exciting source. 
The scales are linear in km s$^{-1}$ and are given by the bars.
\label{fig15}}
\end{figure}
\clearpage

\begin{figure}
\epsscale{1.1}
\plotone{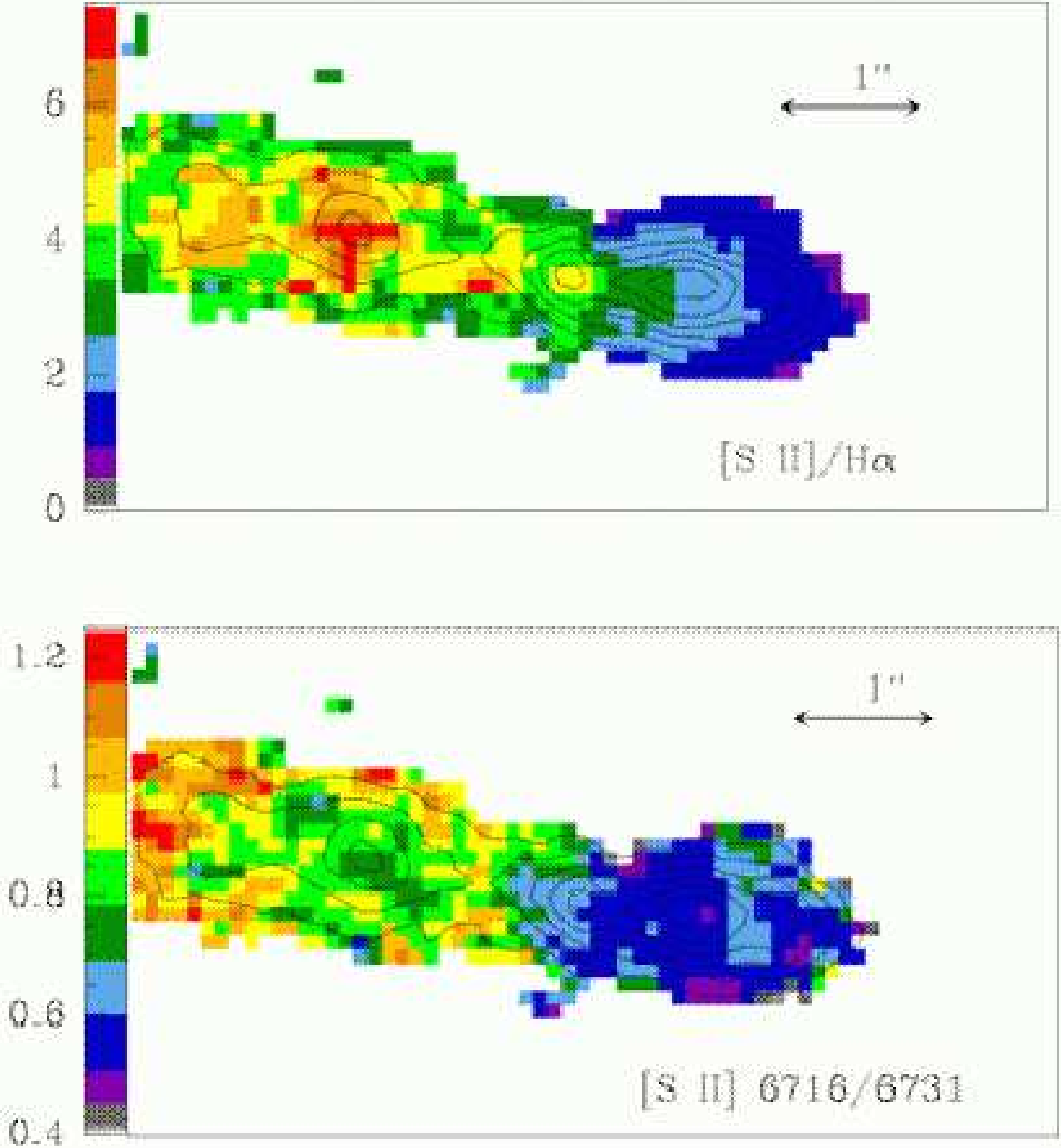}
\caption{[S II] (6716+6731)/H$\alpha$ (upper panel) and [S II] 6716/6731 (bottom panel) ratio maps of the HH 34 exciting source region obtained by summing the [S II] 6716, 6731 \AA\ and H$\alpha$ lines over their wavelength extents in the data cube.
\label{fig16}}
\end{figure}
\clearpage

\begin{figure}
\epsscale{0.7}
\plotone{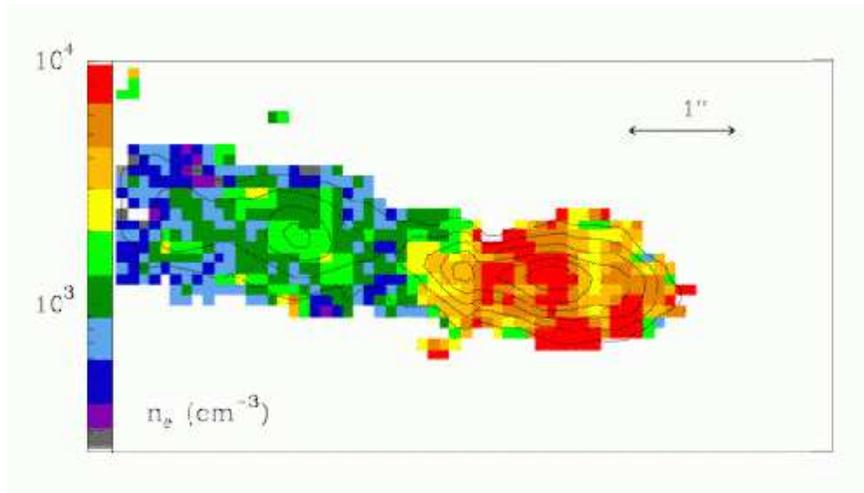}
\caption{The electron density (n$_e$) map obtained from the [S II] 6716/6731 ratio map.  The spatial region corresponds to the HH 34 exciting source.  The scale is logarithmic in cm$^{-3}$ and is shown by the bar.
\label{fig17}}
\end{figure}
\clearpage

\begin{figure}
\epsscale{1.1}
\plotone{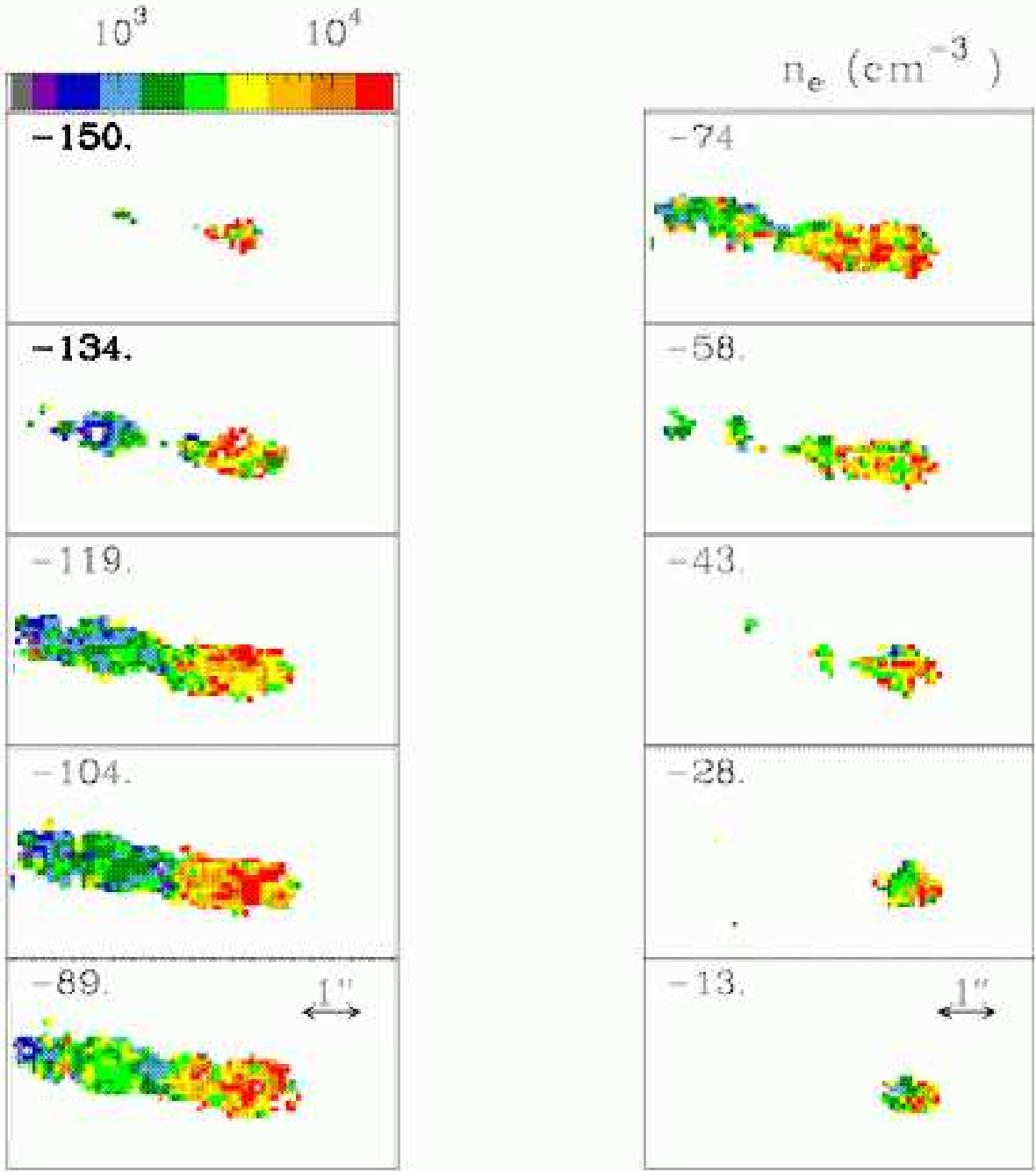}
\caption{The electron density (n$_e$) map obtained from the [S II] 6716/6731 ratio maps 
 at the heliocentric radial velocity shown in each panel.   
The scale is logarithmic in cm$^{-3}$ and is shown by the bar.
\label{fig18}}
\end{figure}
\clearpage








\clearpage

\end{document}